\documentclass[preprint, aps, pra, showpacs]{revtex4-1}
\usepackage{eurosym}
\usepackage{graphicx}
\usepackage{dcolumn}
\usepackage{bm}
\usepackage{amsmath}
\usepackage{xcolor}

\begin{document}

\title{Electron emission perpendicular to polarization direction in laser assisted XUV atomic ionization}

\author{A. A. Gramajo}
\affiliation{Centro At\'omico Bariloche (CNEA) and CONICET, 8400 Bariloche, Argentina}

\author{R. Della Picca}
\affiliation{Centro At\'omico Bariloche (CNEA) and CONICET, 8400 Bariloche, Argentina}



\author{D. G. Arb\'o}

\affiliation{Institute for Astronomy and Space Physics IAFE (UBA-Conicet), Buenos Aires, Argentina}

\date{\today}

\begin{abstract}

We present a theoretical study of ionization of the hydrogen atom due to an XUV pulse in the presence of an IR laser with both fields linearly polarized in the same direction. In particular, we study the energy distribution of photoelectrons emitted perpendicularly to the polarization direction.
By means of a very simple semiclassical model which considers electron trajectories born at different ionization times, the electron energy spectrum can be interpreted as the interplay of \textit{intra-} and \textit{intercycle} interferences. The intracycle interference pattern stems from
the coherent superposition 
of four electron trajectories giving rise to (i) interference of electron trajectories born during
the same half cycle (\textit{intrahalfcycle} interference) and (ii) interference between electron trajectories born during the first half cycle with those born during the second half cycle (\textit{interhalfcycle} interference). The intercycle interference is responsible for the formation of the sidebands. We also show that the destructive interhalfcycle interference for the absorption and emission of an even
number of IR laser photons is responsible for the characteristic sidebands in the perpendicular direction
separated by twice the IR photon energy.
We analyze the dependence of the energy spectrum on the laser intensity and the time delay between the XUV pulse and the IR laser. Finally, we show that our semiclassical simulations are in very good agreement with quantum calculations within the strong field approximation and the numerical solution of the time-dependent Schr\"odinger equation.

\end{abstract}

\maketitle

\preprint{APS/123-QED} 

\section{\label{sec:level1}Introduction}
More than twenty years have passed since the publication of one of the first theoretical predictions of sidebands in laser assisted photoelectric effect (LAPE) \cite{Veniard95}.
The simultaneous absorption of one high-frequency photon and the exchange of several additional photons from the laser field lead to equally spaced sideband peaks in the photoelectron (PE) spectra.
Since this pioneer work, a lot of experiments have been performed in this area. 
Typically, the XUV+ IR field  was firstly obtained through
high-order harmonic generation using the original IR laser field 
as its source \cite{Glover1996,Aseyev2003,OKeeffe2004,Guyetand2005}. 
In contrast to this kind of XUV radiation generation, the monochromaticity of the
femtosecond XUV pulse from a free electron laser (FEL) enables the study of two-color multiphoton ionization without additional overlapping contributions from neighboring harmonics
\cite{Meyer2008, Meyer2010JPB, Radcliffe2012, Mazza2014, Hayden16,Dusterer2016}. 

Several studies have been performed to analyze the PE emission in LAPE depending on different features of the fields: Temporal duration, intensity, polarization state, etc. 
For example, the  temporal overlap between the XUV and IR pulses establishes two well distinguished regimes according to whether the 
XUV pulse duration is greater or less than the laser optical period
\cite{Kienberger2002, Drescher05, Radcliffe2012}.
Whereas in the former, the laser intensity is directly related, in a non-trivial way, to the 
intensity of the appearing sideband peaks in the PE spectrum \cite{Maquet2007,Radcliffe2012,Dusterer2013}, the
latter has been used to characterize the shape and duration of an IR laser pulse with a technique 
called ``streak camera'' 
\cite{Itatani02,Fruehling09,Krausz2009,Pazourek15}.
Furthermore, the variation of the polarization states of each field gives rise to 
dichroic effects in the PE spectrum, which opens the door to the control of the electronic
emission \cite{Taieb2000,Meyer2008,Kazansky2012,Kazansky2014,Mazza2014,Mazza15,Richter2015}.

Hitherto, most of the PE spectra have been measured with angle integrated resolution.
Only very recently it has been possible to measure angularly resolved PE spectra
\cite{Haber2009,Dusterer2013,Mondal2014, Mazza15,Richter2015,Dusterer2016},
which is fundamental to achieve a complete understanding of LAPE process.
In contrast to experiments, theoretical analysis restricted to fixed emission angles
do not present major difficulties.
Most of the theories of LAPE processes are based on the strong field approximation (SFA) 
\cite{Keldysh65,Faisal1973,Reiss1980}.
For example, the soft photon approximation (SPA) \cite{Maquet2007}, derived from the SFA
in the velocity gauge for infinitely long XUV and IR pulses,
depicts satisfactorily the experimental results
\cite{Meyer2006, Meyer2008, Meyer2010JPB, OKeeffe2004, Dusterer2016,Hayden16}.
Besides, the analytic angle-resolved PE spectra derived by Kazansky \textit{et al.} \cite{Kazansky10a,Kazansky10b} and Bivona \textit{et al.} \cite{Bivona10} are based on simplifications of the temporal integration within the SFA.
Furthermore, in our previous work \cite{Gramajo16}, we have presented a semiclassical approach that describes the XUV+IR multiphoton ionization along the direction of polarization of both fields
in very good agreement with the results by solving \textit{ab initio}
the time dependent Schr\"{o}dinger equation (TDSE).
In that work, we have interpreted the PE spectrum as the coherent superposition of electron trajectories emitted with the same optical cycle leading to an intracycle interference pattern
that modulates the sidebands, which can be thought as a consequence of the intercycle interference
of electron trajectories born at different optical cycles. 

To the best of our knowledge, LAPE ionization has not been studied in detail for emission
directions different from the polarization axis.
Furthermore, Haber \textit{et al.} have noted the need for a more comprehensive theoretical description due to the poor agreement between theoretical and experimental PE angular distributions
for the two-color two-photon above threshold ionization \cite{Haber2009,Haber2010}.
Several theories, like SPA, predict no emission in the direction perpendicular to the
polarization axis. However, Bivona \textit{et al.} envisaged non-zero emission for XUV ionization of hydrogen by short intense pulses \cite{Bivona10}. Therefore, in the present paper, we extend
our previous work \cite{Gramajo16} for LAPE from H(1s) to study the emission in the direction perpendicular to the polarization axis of both XUV and laser fields.
In contrast to the case of forward emission, we find that transversal emission
has relatively low probabilities, i.e., the PE energy range is highly reduced.
However, we observe that the PE emission is non-vanishing in agreement with 
Bivona \textit{et al.} \cite{Bivona10}.
Moreover, the PE emission is due to the absorption and emission of an odd number of IR photons
following one XUV photon absorption, whose photoionization line is absent in the PE spectrum. 
Hence, PE spectra in the perpendicular direction can hardly be observed for laser intensities
lower than $10^{13}$ W/cm$^2$. Experimental measurements with strong lasers
would be highly desirable in order to corroborate the present study. A recent work by 
D\"usterer \textit{et al.}  \cite{Dusterer2016} shows that they can be attainable now.

The paper is organized as follows: In Sec. \ref{sec:level2}, we describe the
semiclassical model (SCM) used to calculate the photoelectron spectra for the case of
laser-assisted XUV ionization perpendicular to the polarization direction,
which leads to simple analytical expressions. In
Sec. \ref{results}, we present the results and discuss over the comparison among
of the SCM and the SFA outcomes and the \textit{ab initio} calculation of the TDSE.
Concluding remarks are presented in Sec. \ref{conc}.
Atomic units are used throughout the paper, except when otherwise stated.

\section{\label{sec:level2}Theory of the Semiclassical model}

We study the ionization of an atomic system interacting with an XUV
pulse assisted by an IR laser.
In the single-active-electron (SAE) approximation the TDSE reads
\begin{equation}
i\frac{\partial }{\partial t}\left\vert \psi (t)\right\rangle =
\Big[ H_0  + H_\textrm{int}(t)  \Big]
\left\vert
\psi (t)\right\rangle , 
\label{TDSE}
\end{equation}%
where $H_0=\vec{p}^2/2+V(r)$ is the time-independent atomic Hamiltonian, whose first term
corresponds to the electron kinetic energy, and its second term to the electron-core Coulomb interaction.
The second term in the right-hand side of Eq. (\ref{TDSE}), i.e,  
$H_\textrm{int}=\vec{r} . \vec{F_{X}}(t) + \vec{r} . \vec{F_{L}}(t)$,
describes the interaction of the atom with both time-dependent XUV [$\vec{F_{X}}(t)$]
and IR [$\vec{F_{L}}(t)$] electric fields in the length gauge. 

The electron initially bound in the atomic state $|\phi_{i}\rangle$ is emitted with final momentum $\vec{k}$ and energy $E=k^2/2$ in the final state $|\phi_{f}\rangle$ belonging to the continuum.
Then, the photoelectron momentum distributions can be calculated as
\begin{equation}
\frac{dP}{d\vec{k}}=|T_{if}|^2
\label{prob1}
\end{equation}
where $T_{if}$ is the T-matrix element corresponding to the transition $\phi_{i}\rightarrow\phi_{f}$.

Within the time-dependent distorted wave theory, the transition amplitude in the prior form and length gauge is expressed as
\begin{equation}
T_{if}= -i\int_{-\infty}^{+\infty}dt \,\langle\chi_{f}^{-}(\vec{r},t)|H_\textrm{int}(\vec{r},t)|\phi_{i}(\vec{r},t)\rangle 
\label{Tif}
\end{equation}
where $\phi_{i}(\vec{r},t)=\varphi_{i}(\vec{r})e^{i I_{p} t}$ is the initial atomic state,
$I_{p}$ the ionization potential, and $\chi_{f}^{-}(\vec{r},t)$ is the distorted final state \cite{Macri98,Arbo08a}. 
The SFA neglects the Coulomb distortion in the final channel produced by
the ejected-electron state due to its interaction with the residual ion.
Hence, we can approximate the distorted final state with the Volkov function, which is the solution of
the Schr\"odinger equation 
for a free electron in an electromagnetic field
\cite{Volkov}, i.e.,
$\chi_{f}^{-}= \chi_{f}^{V}$, where
%
\begin{equation}
\chi_{f}^{V}(\vec{r},t)
=(2\pi)^{-3/2}\, \exp{\left[i\big(\vec{k}+\vec{A(t)}\big).\vec{r}+
\frac{i}{2}\int_{t}^{\infty}dt^{'}\big(\vec{k}+\vec{A}(t')\big)^2\right]}
\end{equation}
and the vector potential due to the total external field is defined as 
$\vec{A}(t) = -\int_0^t dt'[\vec{F}_{X}(t')+\vec{F}_{L}(t')]$.

We consider the atomic photoionization due to a short XUV pulse
assisted by an IR laser where both of them are linearly polarized in the same direction $\hat{z}$.
For simplicity, we consider a hydrogen atom initially in the ground state, however,
the present study can be easily generalized to any atom within the SAE approximation.
In the present work, we restrict the photoelectron momentum $\vec{k}=k_z \hat{z} + k_\rho \hat{\rho}$ (in cylindrical coordinates) to the direction perpendicular to the 
polarization axis, i.e., $k_z=0$ and $k_\rho \ge 0$. The case of emission parallel to
the polarization axis, i.e., $k_{\rho} = 0$, was studied recently in \cite{Gramajo16}.

With the appropriate choice of the IR and XUV laser parameters considered, we can assume
that the energy domain of the LAPE processes is well separated from the domain of
ionization by an IR laser alone. 
In other words, the
contribution of IR ionization is negligible in the energy domain where the absorption of one XUV photon takes place. Besides, if we set the general expression of the XUV pulse of duration $\tau_X$
as $\vec{F}_{X}(t)=\hat{z} F_{X0}(t)\cos(\omega _{X}t)$, where $F_{X0}(t)$ is a slowly nonzero varying envelope function, the matrix element can be written as
\begin{equation}
T_{if}= -\frac{i}{2}\int_{t_0}^{t_0+\tau_X}dt\, d_z \big( \vec{k}+\vec{A}(t)\big)\,  F_{X0}(t)\, e^{iS(t)} 
\label{Tif2}
\end{equation}
with $(t_{0},t_{0}+\tau_{X})$ the temporal interval where $F_{X0}(t)$ is nonzero. $S(t)$ is the generalized action 
\begin{equation}
S(t)=-\int_{t}^{\infty}dt'\left[\frac{\big( \vec{k}+\vec{A}(t')\big)^2}{2} + I_{p} -\omega_{X}\right]
\label{action}
\end{equation}
and the $z$-component of the dipole element for the $1s$ state is 
\begin{equation}
d_z(\vec{v})= -\frac{i}{\pi}\, 2^{7/2}(2I_p)^{5/4} 
\frac{ \hat{z}\cdot \vec{v}}{\big[v^{2}+(2I_p)^2\big]^3} .
\label{dip}
\end{equation}
In Eq. (\ref{Tif2}) we have used the rotating wave approximation which accounts, in this case, for the absorption of only one XUV photon and neglects, thus,  
the contribution of XUV photon emission.
As the frequency of the XUV pulse is much higher than the IR laser one, the XUV contribution to the vector potential can be neglected 
\cite{Nagele11, DellaPicca13}, 
regarding that the XUV intensity is not much higher than the laser one.
For the sake of simplicity, we restrict our analysis to the case
where the XUV pulse duration is a multiple of half the IR optical cycle,
i.e., $\tau_{X} = N T_L = 2 N \pi/ \omega_L$, where $T_L$ and $\omega_L$ are the laser period and
the frequency of the IR laser, respectively, and $2N$ is an integer positive number.
During the temporal lapse the XUV pulse is acting,
the IR electric field can be modeled as a cosine-like wave, hence, the
vector potential can be written as $\vec{A}(t)=A_{L0}\sin{(\omega_{L} t)}\hat{z}$
with  $A_{L0}= F_{L0}/\omega_L$ and $F_{L0}$ the amplitude of the laser electric field.

The SCM consists of solving the time integral Eq. (\ref{Tif2}) by
means of the saddle point approximation \cite{Chirila05,Corkum94,Ivanov95,Lewenstein95}, wherein the
transition amplitude can be thought of as a coherent superposition of the amplitudes of all electron classical trajectories with final momentum $\vec{k}$ over the stationary points $t_s$ of the generalized action $S(t)$ in Eq. (\ref{action})
\begin{equation}
T_{if}=\sum_{t_{s}}\frac{\sqrt{2\pi}\, F_{X0}\,d_{z}(\vec{k}+\vec{A}(t_{s}))}
{|\ddot{S}(t_{s})|^{1/2}}\exp{\left[iS(t_{s})+\frac{i\pi}{4}\textrm{sgn}\left(\ddot{S}(t_{s})\right)\right]} ,
\label{Tif3}
\end{equation}
where $\ddot{S}(t) = d^2 S(t)/d t^2 = -\left[ \vec{k}+\vec{A}(t) \right]\cdot \vec{F(t)}$ and $\mathrm{sgn}$ denotes the sign function. Then, from the saddle-point equation, i.e., $\dot{S} = d S(t_s) / d t = 0$, 
the ionization times fulfill the relation
\begin{equation}
A^{2}(t_{s})+k_{\rho}^2 = v_{0}^2  ,
\label{rectats}
\end{equation}
where $v_{0}=\sqrt{2(\omega_{X}-I_{p})}$ is the initial velocity of the electron at the ionization time.
In ionization by an IR laser alone, release times are complex 
due to the fact that the active electron escapes the core via tunneling through the potential
barrier formed by the interaction between the core and the external field, i.e.,
$V(r) + \vec{r}.\vec{F}_L(t)$. Contrarily, in LAPE, real solutions of Eq. (\ref{rectats}) correspond to real ionization
times $t_s$. From Eq. (\ref{rectats}), the domain of allowed classical trajectories perpendicular
to the polarization axis is  $\sqrt{v_{0}^2-A_{L0}^2}  \leq k_{\rho} \leq  v_{0}$
whether $v_0 \ge F_{L0}/\omega_L$.
Non-classical trajectories with complex ionization times have a momentum
$k_{\rho}$ outside the classical domain.
In this work, we neglect the small weight of non-classical trajectories with complex ionization times since its imaginary parts give rise to exponential decay factors.

\begin{figure}[tbp]
\centering
\includegraphics[trim = 1mm 1mm 1mm 1mm, clip, width=0.5\textwidth]{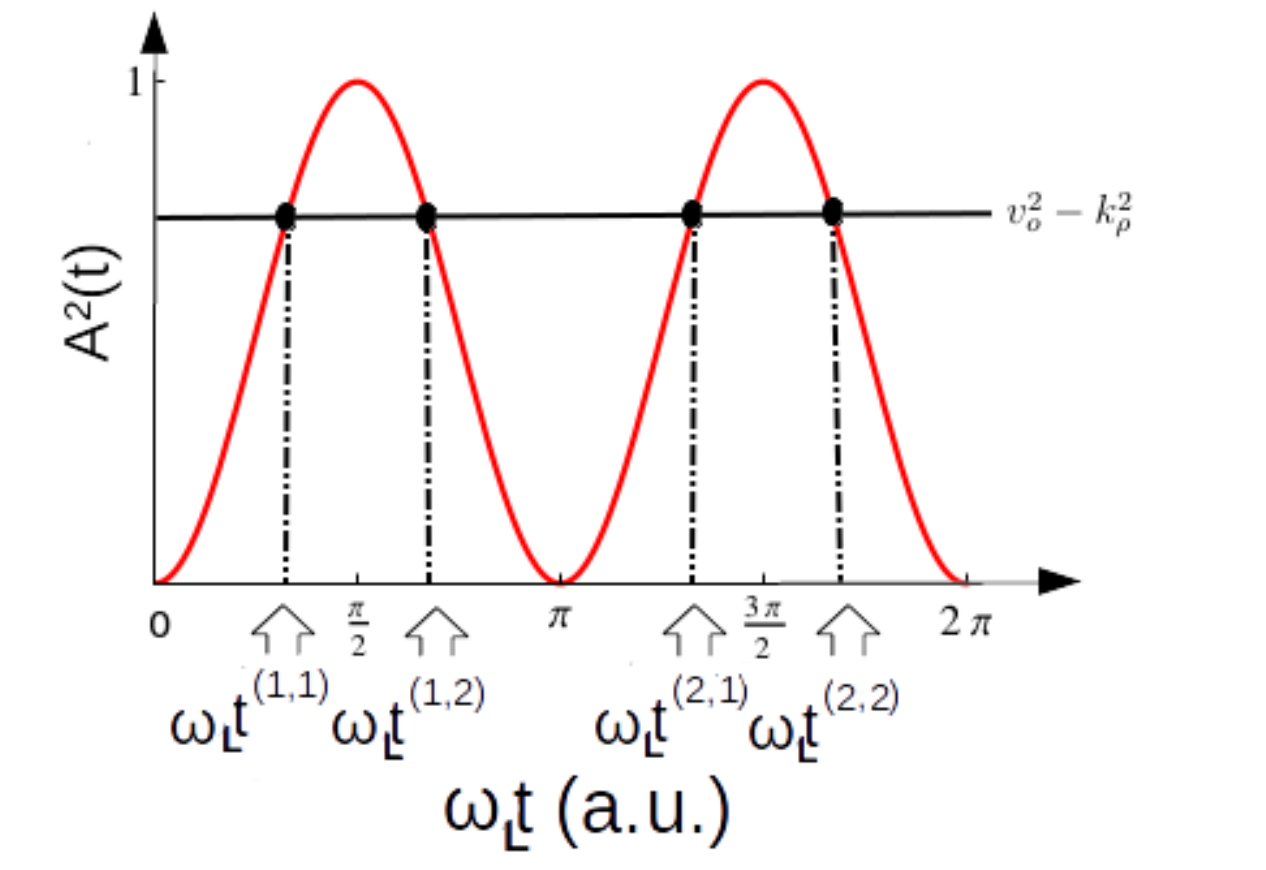}
\caption{Emission times (solutions of Eq. (\ref{rectats})) as the intersection of the two curves 
$A^2(t)=A_{L0}^2\sin^2(\omega_L t)$ in red solid line and $v_0^2-k_\rho^2$ in black solid line for one IR optical cycle. In this particular case, the XUV pulse starts when the potential vector vanishes.}
\label{field}
\end{figure}
The ionization times that verify Eq. (\ref{rectats}) are shown schematically in Fig. \ref{field}
for one IR optical cycle. As we can observe, there are four ionization times per optical cycle and, therefore, the total number of interfering trajectories with the same final momentum perpendicular to the polarization axis is $4N$.
A quick analysis of equations (\ref{Tif3}) and (\ref{rectats}) indicates
that the periodicity for the solution of Eq. (\ref{rectats}) is $\pi/\omega_L$, which is
half of that corresponding to the parallel emission case \cite{Gramajo16}. 
Therefore, the sum over the emission times can be performed alternatively over $2N$ half cycles 
with two emission times in each of them.
They are the early ionization time $t^{(m,1)}$ and the late ionization time $t^{(m,2)}$
corresponding to the $m$-th optical half cycle, where $t^{(m,\beta)}=t^{(1,\beta)} + \pi(m-1)/\omega_{L}$  with $\beta=1,2$. The expressions for the ionization times can be easily derived from Eq. (\ref{rectats}),
\begin{subequations}
\begin{eqnarray}
t^{(1,1)} &=& \frac{1}{\omega_L} \sin^{-1} \Big[\sqrt{(v_0^2-k_\rho^2)/A_{L0}^2} \,  \Big] \label{t11}\\
t^{(1,2)} &=& \frac{\pi}{\omega_L} - t^{(1,1)} . \label{t12}
\end{eqnarray}
\label{t11t12}
\end{subequations}

From Eq. (\ref{action}), the generalized action and its second derivative at the time $t_s$ for electron trajectories along the
perpendicular direction can be written as
\begin{equation}
S(t_s) = \Big( \frac{k_\rho^2}{2} + I_p + U_p -\omega_X \Big) t_s 
- \frac{U_p }{2 \omega_L} \sin(2 \omega_L t_s)
\label{s0}
\end{equation}
and
\begin{equation}
\ddot S(t_s) =  F_{L0}A_{L0} \sin(2 \omega_L t_s)/2 ,
\label{s2}
\end{equation}
respectively, where $U_p=(F_{L0}/2\omega_L)^2$ is the ponderomotive energy of the oscillating electron driven by the laser field.

According to equations (\ref{dip}) and (\ref{t11t12}), the dipole elements $d_z$ evaluated at emission times of consecutive half cycles differ in a sign, i.e.,
\begin{eqnarray}
d_z\big( k_\rho \hat{\rho}+\hat{z}A(t^{(m,\beta)}) \big) & = &  \frac{\sqrt{2} A_{L0} }{i \pi \omega_{X}^{3}} \sin(\omega_L t^{(m,\beta)}) \nonumber \\
& = & - d_z \big( k_\rho \hat{\rho}+\hat{z}A(t^{(m+1,\beta)}) \big) .
\label{dz}
\end{eqnarray}
Hence, the odd and even half cycles have opposite contributions. 
Including equations (\ref{t11t12}) and (\ref{dz}) into Eq. (\ref{Tif3}), the ionization probability of Eq. (\ref{prob1}) can be written as
\begin{eqnarray}
|T_{if}|^2 = \Gamma(k_{\rho})\left|\sum_{m=1}^{2N}\sum_{\beta=1}^{2}(-1)^{m}\exp{\left[iS(t^{(m,\beta)})+\frac{i\pi}{4} \textrm{sgn}\left(\ddot{S}(t^{(m,\beta)}  )\right)\right]}\right|^2 .	
\label{new}
\end{eqnarray}
Eq. (\ref{new}) can be interpreted as the coherent sum 
over interfering trajectories decomposed into those associated with the two release
times within the same half cycle (inner summation)
and those associated with release times in the $2N$ different half cycles (outer summation). 
The ionization probability $\Gamma(k_{\rho})$ contains all identical factors for all subsequent
ionization trajectories which depend on the final momentum $k_{\rho}$, i.e., 
%
\begin{equation}
\Gamma(k_{\rho})= \frac{4 F_{X0}^2 }{\pi \omega_X^6 \omega_L}  
\frac{\sqrt{v_0^2 - k_\rho^2} }{\sqrt{ k_\rho^2 -v_0^2 + A_{L0}^2}} .
\end{equation}

In the same way as in previous works \cite{Gramajo16,Arbo10a,Arbo10b} and after a bit of algebra, it can be shown that
\begin{equation}
\sum_{m=1}^{2N}\sum_{\beta=1}^{2}
(-1)^{m}\,\,e^{\left[iS(t^{(m,\beta)})+\frac{i\pi}{4} \textrm{sgn}\left(\ddot{S}(t^{(m,\beta)} )\right)\right]}
= 2\sum_{m=1}^{2N} e^{i(\bar{S}_{m}+m\pi)}\,\cos{\left(\frac{\Delta{S}_{m}}{2}+\frac{\pi}{4}\right)}
\label{equ}
\end{equation}
where $\bar{S}_{m}=[S(t^{(m,1)})+S(t^{(m,2)})]/2 = S_{0} + m (\tilde{S}/2)$
is the average action of the two trajectories released in the $m$-th half cycle, with $\tilde{S}=(2\pi/ \omega_{L})(E+I_{p}+U_{p}-\omega_{X})$ 
and  $S_{0}$ an unimportant constant that will be canceled out
when the absolute value is taken in Eq. (\ref{new}).
The accumulated action between the two release times $t^{(m,1)}$ and $t^{(m,2)}$ within the same $m$-th half cycle, $\Delta{S}_{m}=S(t^{(m,1)})-S(t^{(m,2)})$ in Eq. (\ref{equ}), is given by
\begin{eqnarray}
\Delta{S}&=&\frac{\tilde{S}}{2} \left  \{\frac{2}{\pi} \sin^{-1}
\left[\sqrt{(v_0^2-k_\rho^2)/A_{L0}^2} \right] -  1 \right \}  \nonumber\\
&&- \frac{1}{2\omega_{L}}\sqrt{v_0^2-k_{\rho}^2}\sqrt{ k_{\rho}^2-v_{0}^2+A_{L0}^2},
\label{DeltaS}
\end{eqnarray}
where we have omitted the subscript $m$, since it is independent of which half-cycle 
is considered. Finally, due to the linear dependence of the average action $\bar{S}_{m}$
on the cycle number $m$ and the factorization of the cosine factor in the right side of Eq. (\ref{equ}),
the ionization probability can be easily written as
\begin{subequations}
\begin{eqnarray}
|T_{if}|^2&=&4\Gamma{(k_{\rho})}\underbrace{\cos^{2}{\left(\frac{\Delta{S}}{2} +\frac{\pi}{4} \right)}}_{\textrm{intrahalfcycle}} \underbrace{\left[\frac{\sin{(N\tilde{S}/2)}}{\cos{(\tilde{S}/4)}}\right]^2}_{\textrm{interhalfcycle}} .
\label{probb}  \\
&=&4\Gamma{(k_{\rho})}\underbrace{4\cos^{2}{\left(\frac{\Delta{S}}{2} +\frac{\pi}{4} \right)}
\sin^2\left(\frac{\tilde{S}}{4}\right)}_{\textrm{intracycle}} \underbrace{\left[\frac{\sin{(N\tilde{S}/2)}}{\sin{(\tilde{S}/2)}}\right]^2}_{\textrm{intercycle}}
\label{proba}
\end{eqnarray}
\label{prob}
\end{subequations}

Equations (\ref{probb}) and (\ref{proba}) indicate that the photoelectron spectrum can be factorized in two different ways. On one hand, (i) the factorization in Eq. (\ref{probb}) highlights the 
contribution of the pair of electron trajectories within the same half cycle 
(\textit{intrahalfcycle} interference), governed by the factor 
$G(k_{\rho})=\cos^{2}\left(\Delta{S}/2 + \pi/4 \right)$,
and the interference stemming from trajectories released at different half cycles (\textit{interhalfcycle} interference) described by the factor
$H(k_{\rho})=\left[ \sin{(N\tilde{S}/2)}/ \cos{(\tilde{S}/4)} \right]^2$.
On the other hand, (ii) the factor 
$F(k_{\rho})=4\cos^{2}\left(\Delta{S}/2 + \pi /4 \right)\sin^2(\tilde{S}/4)$
stemming from the contribution of the four trajectories within the same optical
cycle (\textit{intracycle} interference), and the factor 
$B(k_{\rho})=\sin^2(N \tilde{S}/2)/\sin^2(\tilde{S}/2)$
stemming from trajectories released at different cycles (\textit{intercycle} interference, in correspondence with previous analysis of Eq. (23) in \cite{Gramajo16}).
Whereas in (i) the interference of $2 N$ half
cycles is highlighted giving rise to the intrahalf- and interhalfcycle factors,
in (ii) we think of the coherent contributions of
$N$ different optical cycles splitting the contribution in intra- and intercycle interference patterns.
Obviously, the two different factorizations give rise to the same results, i.e., 
$G(k_{\rho})H(k_{\rho}) = F(k_{\rho})B(k_{\rho})$.

\begin{figure}[tbp]
\centering
\includegraphics[trim = 1mm 8mm 1mm 1mm, clip, width=0.6\textwidth]{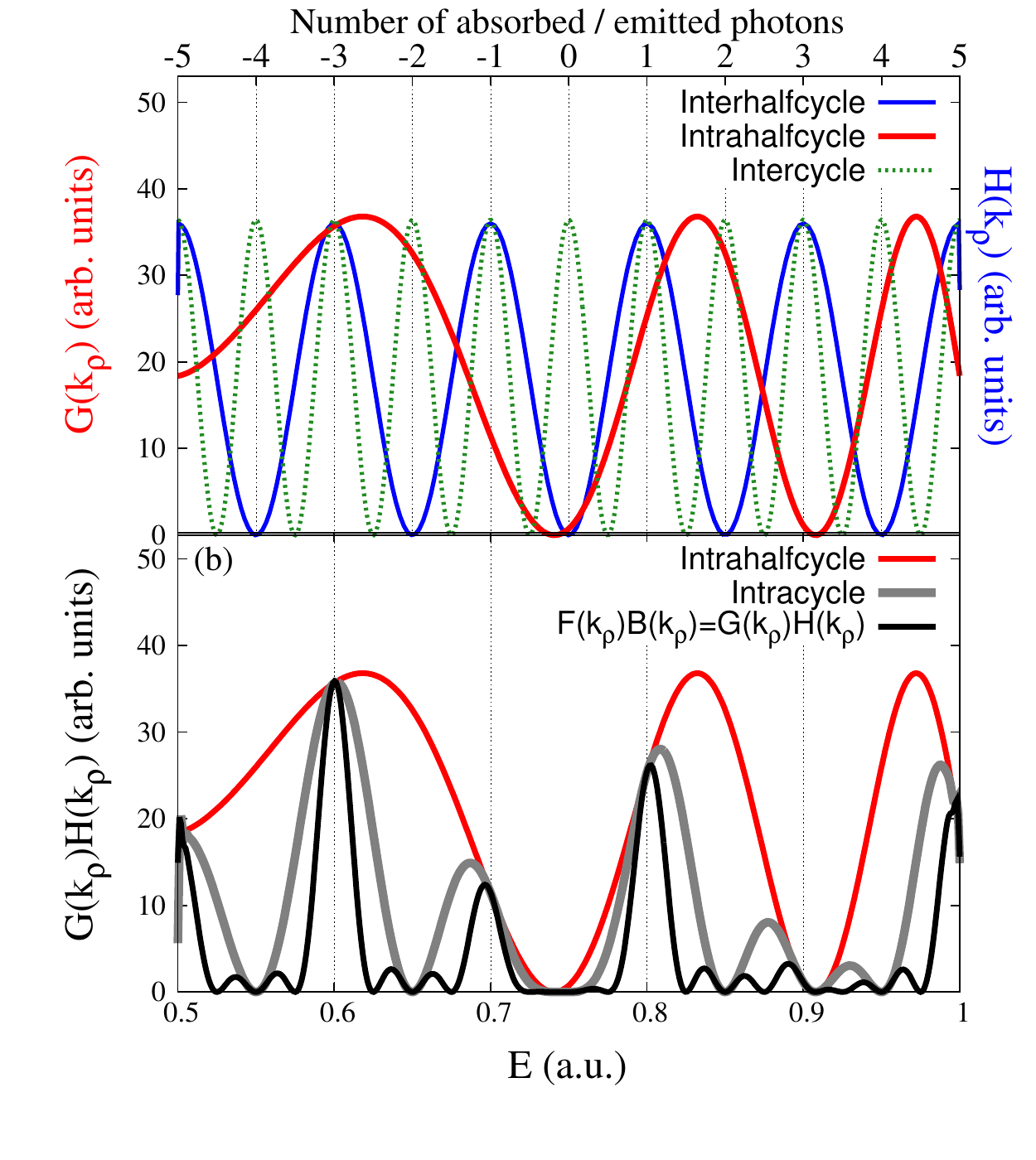}
\caption{Buildup of the interference pattern following the SCM for $N=2$.
(a) Intrahalfcycle interference pattern given by $G(k_{\rho})$ in red line, intercycle
pattern given by the factor $B(k_{\rho})$ in 
green dotted line and interhalfcycle pattern given by the factor $H(k_{\rho})$ in blue line.  
(b) Intracycle pattern given by the factor $F(k_{\rho})$ in grey line and total interference pattern $F(k_{\rho})B(k_{\rho})=G(k_{\rho})H(k_{\rho})$ in black line. Vertical lines depict the positions of the SBs $E_{\ell}$ of Eq. (\ref{El}). The IR laser parameters are $F_{L0}=0.05$, $\omega_{L}=0.05$, and $\tau_{L}=5T_{L}$ and the XUV frecuency is $\omega_{X}=1.5$.}
\label{FBGH}
\end{figure}

In Fig. \ref{FBGH}(a), we plot the intrahalfcycle function $G(k_{\rho})$
and the interhalfcycle $H(k_{\rho})$ for a XUV laser pulse duration $\tau_{X}=2T_{L}$
as a function of the energy.
Whereas the intrahalfcycle factor $G(k_{\rho})$ exhibits a non-periodic
oscillation, the interhalfcycle $H(k_{\rho})$ is periodic in the final 
photoelectron energy with peaks at positions $E_{\ell} = k_{\rho}^2/2$
given by 
\begin{equation}
E_{\ell} = \omega_X + (2\ell+1) \omega_L  - I_p - U_p .
\label{El}
\end{equation}
with $\ell = ...,-2,-1,0,1,2,...$. In fact, in the limit of infinitely long XUV and IR pulses, $\lim_{N\rightarrow \infty}H(k_{\rho})=\sum_{l}\delta(\tilde{S}/4 + \pi/2 - \ell \pi)$, the ionization probability vanishes unless the final energy satisfies Eq. (\ref{El}) which gives the positions of the different sidebands.
We see that the energy difference between two consecutive sidebands is
$2\omega_L$ and not $\omega_L$ as for the emission in the direction
parallel to the polarization axis \cite{Gramajo16}. From Eq. (\ref{El}), it is easy to see that
only odd numbers ($2\ell + 1$) of laser photons can be absorbed or emitted together with the 
absorption of one XUV photon. Due to the lack of sidebands for the absorption or emission
of an even number of laser photons, the absorption of only one XUV photon alone (in the absence
of absorption or emission of IR photons) is forbidden.
The intrahalfcycle pattern displays few oscillations with maxima depending
on the electron kinetic energy. These can be easily calculated through $\Delta{S}=(2q-1/2)\pi$, with integer $q$. 
In Fig. \ref{FBGH}(b), we plot the total interference pattern corresponding to
an XUV pulse of duration $\tau_{X}=2T_{L}$, and the intracycle factor $F(k_{\rho})$.
For the sake of comparison, we reproduce in Fig \ref{FBGH}(b) the intrahalfcycle factor $G(k_{\rho})$ 
of Fig \ref{FBGH}(a).
The multiplication of both \textit{intrahalfcycle} and \textit{interhalfcycle} factors,
i.e., $G(k_{\rho})H(k_{\rho})$, is displayed in Fig. \ref{FBGH}(b),
where we observed how the intrahalfcycle interference pattern [$G(k_{\rho})$] works
as a modulation of the intracycle interference pattern [$F(k_{\rho})$] and the
latter does the same with the sidebands (intercycle interference pattern).

On the other hand, Eq. (\ref{proba}) shows that the photoelectron spectrum
can be thought as the intercycle pattern with peaks at positions 
$E_n = n \omega_L + \omega_X - I_p - U_p$ modulated by the intracycle interference
pattern given by the factor $F(k_{\rho})$. Therefore, the lack of even-order sidebands
stems from the factor $\sin^2(\tilde{S}/4)$ into the intracycle factor
$F(k_{\rho})$ [see Eq. (\ref{probb})].
The factor $\sin^2(\tilde{S}/4)$ reflects the fact that
the dipole element has opposite signs for the two different half cycles
into the same optical cycle [see Eq. (\ref{dz})] giving rise to destructive
interference between the contribution of the two electron trajectories of the
first half cycle with the corresponding to the second half cycle of every optical cycle
during the time interval that the XUV pulse is on. Contrarily,
for emissions in the parallel direction, whereas the ionization during
one of the two half cycles contributes to emissions in one direction (forward or backward),
the other half cycle will contribute to the opposite direction \cite{Gramajo16}.
Therefore, no interference is produced for parallel emissions
allowing to all peaks separated by $\omega_L$.


\section{Results and discussion}
\label{results}

At the time of probing the general conclusion of the SCM that the
ionization probability of electrons emitted perpendicularly to the
polarization axis of the XUV and the laser pulse can be factorized in two
different contributions in two different ways:
(i) \textit{intrahalfcycle} and \textit{interhalfcycle} interferences [Eq. (\ref{probb})] and
(ii) \textit{intracycle} and \textit{intercycle} interferences [Eq. (\ref{proba})],
we need to compare the outcome of SCM calculations with quantum ones.
We have performed calculations within the SFA and TDSE methods, which have been
extensively covered in the literature 
and in our previous work \cite{Gramajo16}
and we do not repeat here.
For the SFA calculating method, please refer, for example, to Refs.
\cite{Arbo08a,Kazansky10a,Kazansky10b,Bivona10,Arbo10a,Arbo10b}, and for the 
\textit{ab initio} numerical solutions of the TDSE we employ the 
generalized pseudospectral method combined with the split-operator representation 
of the time-evolution operator, which is thoroughly explained in the literature (see,
for example, \cite{Tong97,Tong00,Tong05}).
For the computational feasibility of the SFA and TDSE calculations, 
we take the XUV pulse and the IR laser field modeled as 
\begin{equation}
\vec{F}_{i}(t)=F_{i0}(t-t_{ib})\ \cos \left[ \omega _{i}\left( t-
\Delta_i-\frac{\tau_{L}}{2}\right) \right] \ \hat{z},
\label{i-field}
\end{equation}%
where $i=$L and X denote the IR laser and XUV pulses, respectively. 
The envelopes of the electric fields in Eq. (\ref{i-field})
were chosen as
%
\begin{eqnarray}
F_{i0}(t)=F_{i0}\left\{ 
\begin{array}{ccc}
t/T_i           & \mathrm{if} & 0   \leq t\leq T_i \\ 
1               & \mathrm{if} & T_i \leq t\leq \tau_i - T_i \\ 
(\tau_{i}-t)/T_i& \mathrm{if} & \tau_i - T_i \leq t \leq \tau_i \\
\end{array}
\right.
\label{i-envelope}
\end{eqnarray}
and zero otherwise, where $T_i= 2\pi/\omega_i$ and $\tau_i$ are the $i-$field period and pulse duration,
respectively. It describes a central flattop region and linear one-cycle ramp on and ramp off.
For the sake of simplicity, we suppose that the duration of both
laser fields comprise integer number of cycles, i.e., $\tau_i = N_i T_i$ where $N_i$ is a
positive integer.
In addition, as we have mentioned before, we also consider the case where $\tau_X = N T_L$.
We choose the origin of the time scale as the beginning of the IR laser pulse,
i.e., $t_{Lb}=0$, with no displacement of the laser pulse $\Delta_L=0$.
In this way, the IR laser field is a cosine-like pulse centered in the middle of 
the pulse, $t=\tau _{L}/2$.
In Eq. (\ref{i-field}), the time delay of the XUV pulse with respect to the laser
pulse is $\Delta_X$ and $t_{Xb}= \Delta_X + \tau_L/2 - \tau_X/2$ denotes the beginning of the XUV pulse
that depends on the XUV pulse duration. It also marks the starting time of 
the active window for ionization.
The vector potential from the perspective of the active window is shifted when
comparing different XUV pulse durations due to the different values of $t_{Xb}$.
Therefore, for the sake of comparison of the ionization yield for different XUV pulse durations,
the active window should be in phase with the vector potential. 
For that, we define the module $2 \pi$ optical phase
$\phi \equiv  \omega_L t_{Xb}  = \omega_L \Delta_X + (N_L-N) \pi$
as the phase of the starting time of the XUV pulse with respect to the vector potential $\vec{A}(t)$
\footnote{Here the $2\pi$ equivalence  $a \equiv b $ means that $(a -b)/2\pi$ is integer and $0\leq \phi < 2\pi$.}.

\begin{figure}
\centering
\includegraphics[trim = 1mm 8mm 1mm 5mm, clip, width=0.65\textwidth]{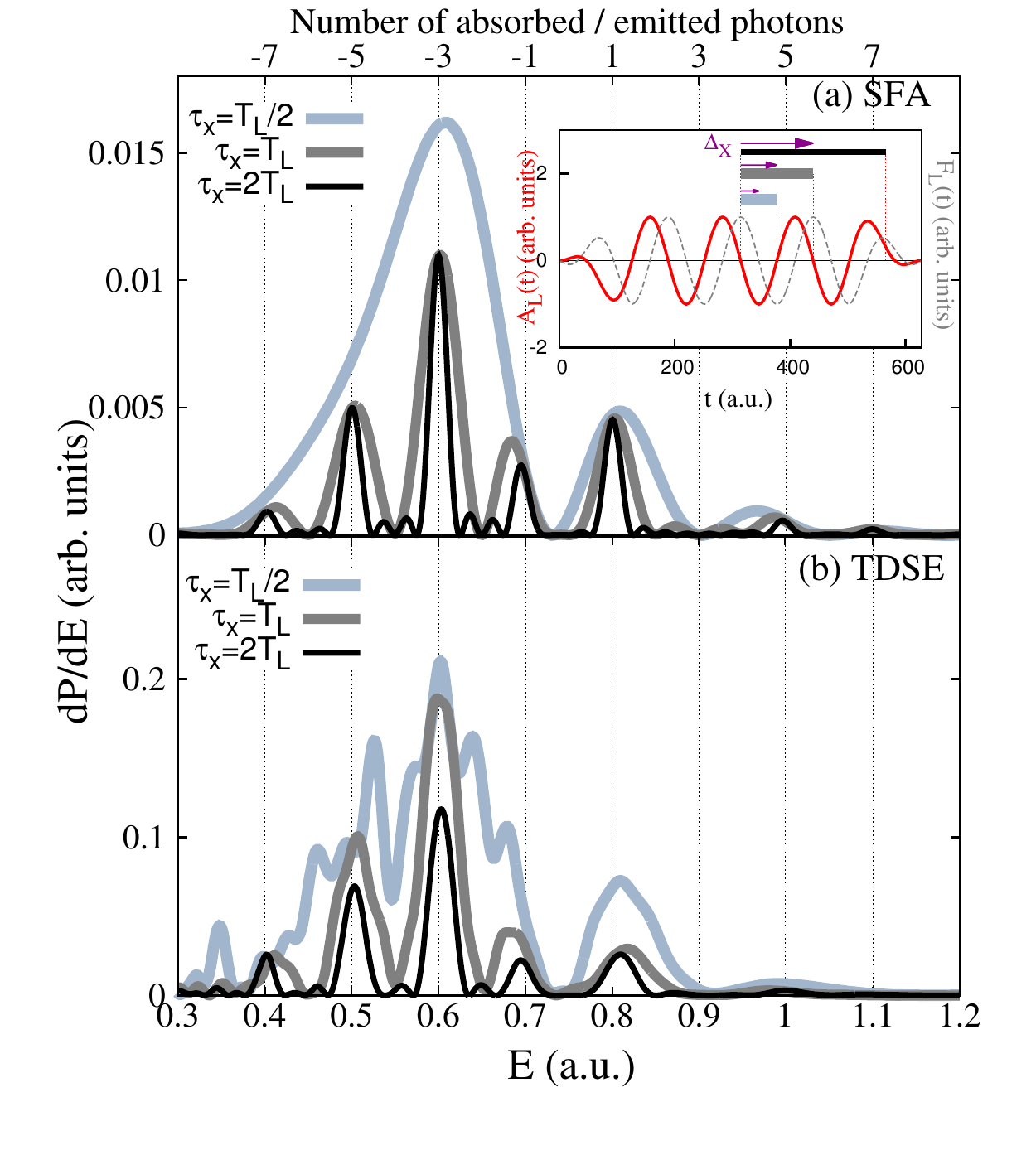}
\caption{Photoelectron spectra in the perpendicular direction calculated within (a) the SFA and 
(b) the TDSE, for different XUV pulse durations $\tau_{X}=T_{L}/2, T_{L}$, and $2 T_{L}$ and 
respective time delays $\Delta_X= T_L/4$, $T_L/2$, and $T_L$.
The XUV and IR  parameters are the same as in Fig. \ref{FBGH} and $F_{X0}=0.05$. Vertical lines
depict the positions of the sidebands according to Eq. (\ref{El}).}
\label{comparison}
\end{figure}

\begin{figure}[h!]
\centering
\includegraphics[angle = 0, trim = 1mm 2mm 1mm 5mm, clip, width=0.8\textwidth]{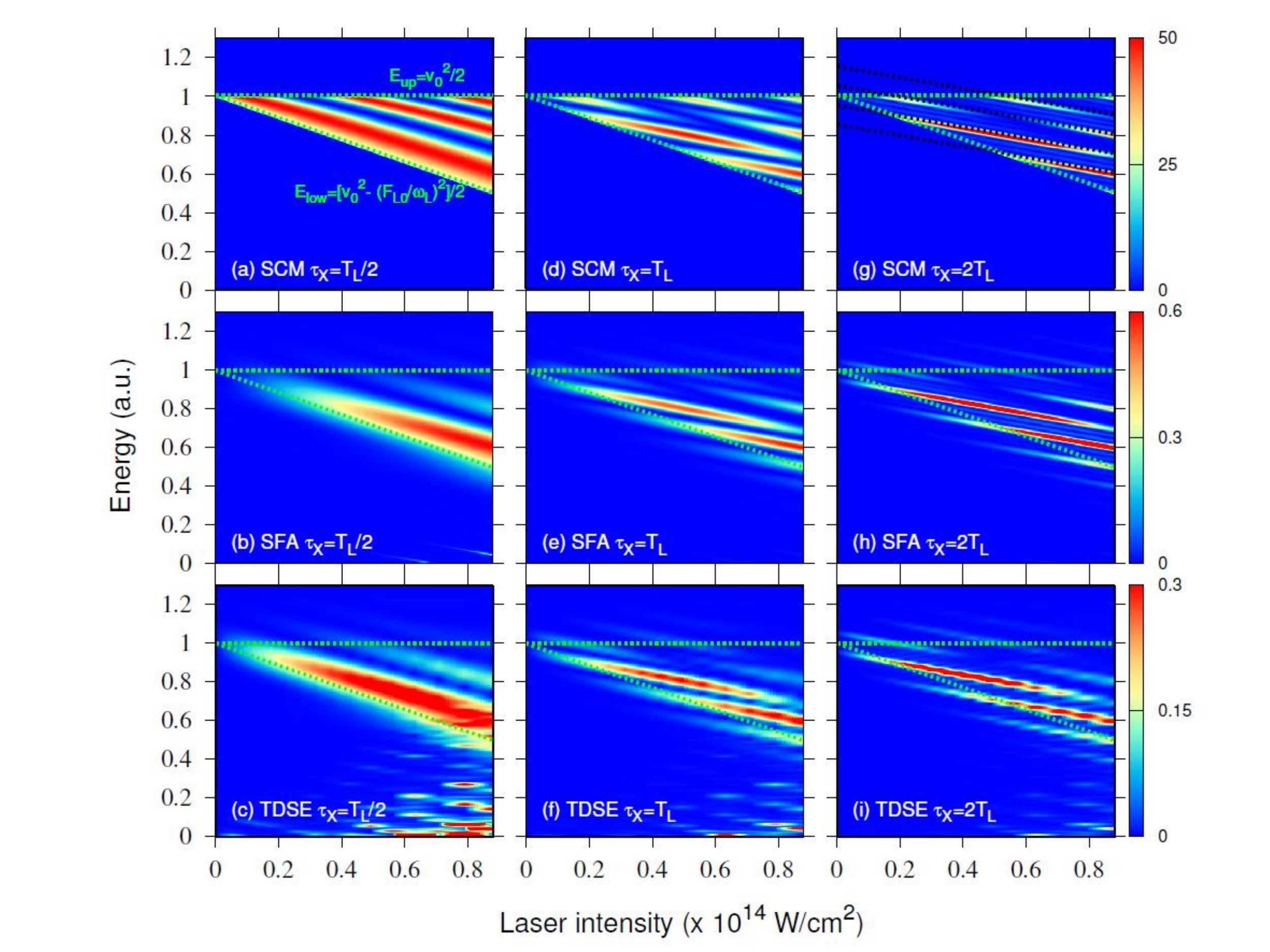}
\caption{Photoelectron spectra in the perpendicular direction (in arbitrary units) calculated at
different laser field strengths within the SCM (a, d, and g), the SFA (b, e, and h),
and the TDSE (c, f, and i). The XUV pulse durations are $\tau_{X}=T_L/2$ (a-c),
$\tau_{X}=T_L$ (d-f), and $\tau_{X}=2 T_L$ (g-i).
The other XUV and IR parameters as in previous figures. 
In green dotted line we show
the classical boundaries and in black dotted line the $E_\ell$ values given by Eq. (\ref{El}).}
\label{EvsI}
\end{figure}

In the following, we probe the results of the SCM by comparing them to quantum simulations. We consider the IR and XUV frequencies as $\omega_L=0.05$ and $\omega_X=30 \omega_L = 1.5$ respectively, the IR laser duration $\tau_L = 5 T_L$ and three different XUV pulses with durations $\tau _{X}=T_{L}/2$, $T_{L}$, and $2T_{L}$ (\textit{i.e.} $N=1/2$, 1 and 2).
In Figs. \ref{comparison} and  \ref{EvsI} we consider the corresponding time delays $\Delta_{X}= T_L/4$, $T_L/2$, and $T_L$, so that the optical phases are the same $\phi=\pi$.
In Fig. \ref{comparison}(a) and \ref{comparison}(b) we show results of the SFA and 
the numerical solution of the TDSE, respectively,
for the same XUV and IR pulse parameters used in Fig. \ref{FBGH} with $F_{X0}=F_{L0}=0.05$.
The agreement among the SCM [Fig. \ref{FBGH}], the SFA [Fig. \ref{comparison}(a)],
and TDSE [Fig. \ref{comparison}(b)] energy distributions is very good since
the effect of the Coulomb potential on the energy spectrum
for electron emission in the perpendicular direction is very small if not negligible.
However, the analysis
of the effect of the Coulomb potential of the remaining core on the electron yield 
deserves a thorough study, which is beyond the scope of this paper.
As predicted in Eq. (\ref{probb}), the intrahalfcycle interference
pattern, calculated as the energy distribution for a XUV pulse duration of half a laser cycle,
i.e., $\tau_{X} = T_{L}/2$, modulates the intracycle interference
pattern, calculated as the energy distribution for a XUV pulse duration of one laser cycle,
i.e., $\tau_{X} = T_{L}$. In the same way, the latter modulates the sidebands in the energy
distribution for a longer XUV pulse, i.e., $\tau_{X} = 2 T_{L}$, as shown 
in Fig. \ref {comparison}(a) and Fig. \ref {comparison}(b). For the latter case,
(when the XUV pulse duration involves several periods of the laser, i.e., $\tau_{X}=2T_{L}$),
the positions of the sidebands obtained by the quantum calculations (SFA and TDSE) in Figs. \ref{comparison}(a) and \ref{comparison}(b) agree with the SCM expressed in Eq. (\ref{El}). 
As expected, the energy spectra for the quantum SFA and TDSE calculations extend
beyond the classical limits 
$E_{\textrm{low}}=v^{2}_{0}/2 - 2 U_p = 0.5$ and $E_{\textrm{up}}= v^{2}_{0}/2 = 1$.

The TDSE spectrum for the shorter XUV duration case in Fig. \ref{comparison}(b), present several additional structures that are related to the direct electronic emission due to the IR laser only. This is discussed in the context of the next figure.

We have investigated over the dependence of the energy distribution for photoelectrons emitted in the direction perpendicular to the polarization axis on the intensity of the XUV pulse. We have checked that the total (angle- and energy-integrated) ionization probability is essentially proportional to the 
intensity of the XUV pulse whereas the overall shape of the energy distribution
in the transversal direction (Fig. \ref{comparison}) remains rather
unchanged when varying the intensity of the XUV pulse (not shown). Contrarily, 
the intensity of the IR laser has a strong effect on the shape of the energy distribution.
In Fig. \ref{EvsI} we show calculations of the
energy distribution in the perpendicular direction within the SCM (a), (d), and (g),
the SFA in  (b), (e), and (h), and the TDSE in  (c), (f), and (i),
for laser field intensities from $I_{L}=0$ up to $8.8 \times 10^{13}$ W/cm$^2$ ($F_{L0}=0.05$).
We analyze the energy distribution for different XUV pulse durations.
The energy spectra for $\tau_X = T_L /2$, $T_L$, and $2T_L$ in Figs. \ref{FBGH} and
\ref{comparison} are cuts of Fig. \ref{EvsI} at $I_{L}=8.8 \times 10^{13}$ W/cm$^2$.
The classical boundaries $E_{\textrm{low}}$ and $E_{\textrm{up}}$ drawn in dotted lines
exactly delimit the SCM spectrogram of Fig. \ref{EvsI}(a), (d) and (g), as expected.
For the case where $\tau_X = T_L/2$ (first column), Figs. \ref{EvsI} (a), (b), and (c)
show a negative slope of the intrahalfcycle interference stripes.
The value of the slope
for the maxima can be calculated numerically from the transcendental equation for the energy
$\Delta S = (2q-1/2)\pi$ with $q=-2,-1,1,2,...$ [see Eq. (\ref{DeltaS})].
For the cases $\tau_X = T_L$ (second column of Fig. \ref{EvsI}),
we observed in Figs. \ref{EvsI}(d), (e), and (f)
that the intrahalfcycle interference patterns are flanked by stripes of zero or near zero probability distribution corresponding to the zeros of the factor $\sin(\tilde{S}/4)$ in the intracycle factor $F(k_\rho)$,
i.e., $\tilde{S}/4 = n \pi$, which gives $E = \omega_X - I_p - U_p + 2 n \omega_L$.
The slope of these minima is $-U_p/I_L=-(2 \omega_L)^{-2}$ and the 
energy difference between consecutive minima (and maxima) is $2 \omega_L$. 
For the case of $\tau_X = 2 T_L$ (third column of Fig. \ref{EvsI}),
we see in Fig. \ref{EvsI} (g), (h), and (i) that the stripes of the probability distribution become even thinner 
due to the effect of the destructive intercycle interference
for energy values much different from the conservation energy for absorption of one XUV photon and
an odd number of IR laser photons [Eq. (\ref{El})]. Moreover, when we compare the position of the maxima with Eq. (\ref{El}),
marked as black dotted lines in Fig. \ref{EvsI} (g),
we see an excellent agreement (see also Fig. \ref{comparison}). 
The domain of the SFA and TDSE spectrograms (second and third row of Fig. \ref{comparison})
extend beyond the classical boundaries with smooth edges.
The characteristic intrahalf- and intracycle stripes with negative slope
reproduce very well the SCM predictions.
In Figs. \ref{EvsI} (c), (f), and (i), the TDSE calculations exhibit a strong
probability distribution for high values of the laser intensity 
$I_{L} \gtrsim 0.5 \times 10^{14}$ W/cm$^2$ in the low energy region which 
almost does not overlap with the laser assisted XUV ionization
for the longer  XUV duration cases, but for the $\tau_X=T_L/2$ case.
We suspect that these structures are responsible of those appearing in Fig \ref{comparison}(b). 
The source of this probability enhancement is the atomic ionization by the IR laser pulse alone,
which has not been considered in our SCM and is strongly suppressed in the SFA
because the laser photon energy is much lower than the ionization potential,
i.e., $\omega_{L} \ll I_{p}$. 
For this reason, we can confirm that the SFA is a more reliable method to deal with laser assisted photoemission copared to ATI by IR lasers \cite{Gramajo16}. Therefore, except for the region where ionization by the laser field alone becomes important, SFA and TDSE spectrograms exhibit a very good agreement between them
and resemble the SCM calculations qualitatively well.
The resulting energy stripes become thinner and more pronounced as the duration of the XUV pulse
increases, exhibiting the fact that the intrahalfcycle interference pattern 
modulates the intracycle pattern, which, at the same time, modulates the sidebands 
(intercycle interference pattern).

\begin{figure}[h!]
\centering
\includegraphics[angle = 0, trim = 1mm 1mm 1mm 5mm, clip, width=0.8\textwidth]{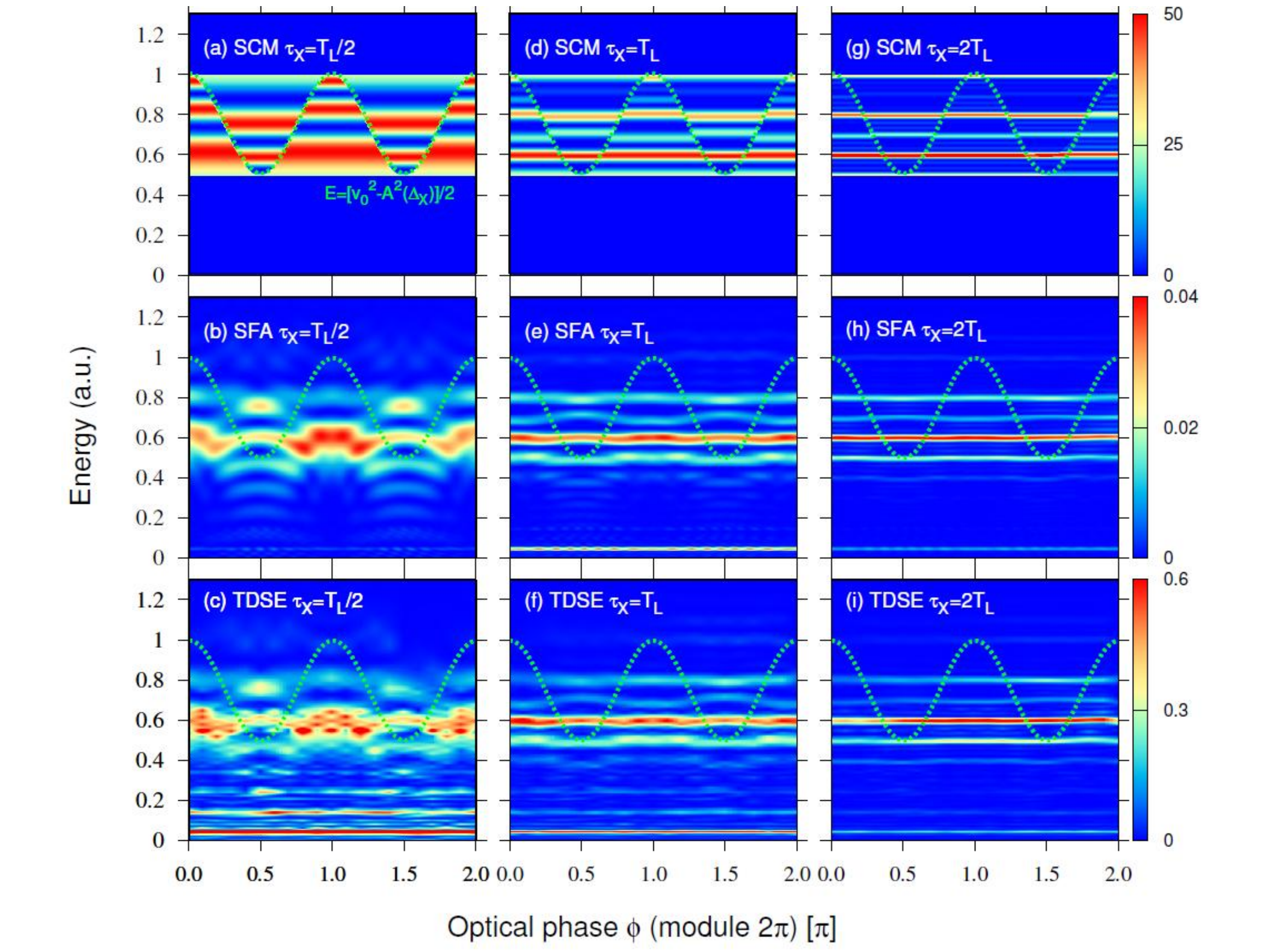}
\caption{Photoelectron spectra in the perpendicular direction (in arbitrary units) calculated at
as a function of the time delay $\Delta_X$ within the SCM [(a), (d), and (g)],
the SFA [(b), (e), and (h)], and the TDSE [(c), (f), and (i)].
The XUV pulse durations are $\tau_{X}=T_L/2$ [(a)-(c)],
$\tau_{X}=T_L$ [(d)-(f)], and $\tau_{X}=2 T_L$ [(g)-(i)].
The other XUV and IR  parameters as in previous figures.
In dotted line we show the 
energy values of Eq. (\ref{Edis}).
}
\label{Evst12}
\end{figure}
So far, we have performed our analysis of the electron emission in the
transverse direction for optical phase $\phi=\pi$ (since $N_L$ is odd).
In order to reveal how the intracycle interference pattern changes with the time delay,
we vary $\Delta_X$ in an optical cycle, so that $\phi$ varies from $0$ to $2\pi$. 
In Fig. \ref{Evst12}(a) we show the intrahalfcycle interference pattern calculated for 
$\tau_X = T_L /2$ within the SCM in the transverse direction as a function
of the optical phase module $2\pi$.
The horizontal stripes show the independence of the
intracycle interference pattern with the time delay, except for the
discontinuity for energy values equal to
\begin{equation} 
E_{\mathrm{disc}}=\frac{1}{2} \left[ v_{0}^2-A^2(t_{Xb})\right].
\label{Edis}
\end{equation}
For $\phi=0$ the discontinuity is situated at $E_{\mathrm{disc}}=v_{0}^{2}/2$
[since in this case $\Delta_X = 3 T_L/4$ for $N=1/2$ and then $A(t_{Xb}=3 T_L)=0$],
which coincides with the classical boundary.
Fig. \ref{Evst12}(a) shows us that as $\phi$ (and $\Delta_X$) varies,
the discontinuity follows the shape of the square of the vector potential,
which means that the discontinuity is $\pi$-periodic in $\phi$, contrarily to the
$2\pi-$periodicity in the case of parallel emission \cite{Gramajo16}.
For phase values $ \phi=0, \pi,$ and $2\pi$, the discontinuity situates
at $E_{\textrm{up}}=v_0^2/2=1$, whereas for $\phi= \pi/2$, and $3\pi/2$, it does at 
$E_{\textrm{low}}=v_0^2/2-2U_p=0.5$, losing entity in both cases.
The SFA and TDSE energy distributions, in the respective
Fig. \ref{Evst12}(b) and (c), exhibit similar characteristics to the SCM,
but with a richer $\pi-$periodic structure. Interestingly, the discontinuity at
$E_{\mathrm{disc}}$ is reflected as a jump of the probability distributions for the same energy values.
The remarkable resemblance between the computationally cheap SFA and the \textit{ab initio} solution of the TDSE results shows, once again, 
that the SFA is very appropriate to explain and reproduce the electron yield in LAPE processes.
Low energy contributions in TDSE calculations shown in Fig. \ref{Evst12}(a)
are due to IR ionization as described before in Fig. \ref{EvsI}(c), (d) and (e).

For $\tau_X = T_L$ in Fig. \ref{Evst12}(d), the SCM spectrum displays horizontal lines
corresponding to the intracycle interference or, what is the same,
to the interplay between the intrahalfcycle factor $G(k_{\rho})$ and the
factor $\sin^2(\tilde{S}/4)$, according to Eq. (\ref{proba}). In the same way,
for the case where $\tau_X = 2 T_L$ in Fig. \ref{Evst12}(g) the SCM spectrum displays horizontal lines
corresponding to the intercycle interference modulated by the intracycle pattern of 
Fig. \ref{Evst12}(d). We note that there is no discontinuity in factor $G(k_{\rho})$
at the energy values $E_\ell$ given by Eq. (\ref{El}). 
Hence, as the sidebands get narrower, discontinuity of the intracycle modulation blurs. 
Continuity in the intra- and intrahalfcycle factors is related to the fact that 
the accumulated action at both sides of the discontinuity verifies that  
$\Delta S\vert_{E>E_\textrm{disc}}+ \Delta S\vert_{E<E_\textrm{disc}} = \tilde{S}/2$,
where $\Delta S\vert_{E>E_\textrm{disc}}$ ($\Delta S\vert_{E<E_\textrm{disc}}$)
is the accumulated action calculated at energies higher (lower) than the discontinuity $E_\textrm{disc}$. Hence, the evaluation of $\cos^2(\Delta S/2 + \pi/4)$
gives exactly the same result at $E_\ell$ independently on $\phi$. 
Once more, from the SFA spectrograms displayed in Fig. \ref{Evst12}(e) and (h) and 
the corresponding TDSE calculations in Fig. \ref{Evst12}(f) and (i), we can see,
once again, that the agreement between the SFA and TDSE spectrograms is very good,
with the exception of a contribution at low energies due to
the ionization by the IR laser pulse alone, which is strongly suppressed in the
SFA calculations. By comparing the intrahalfcycle pattern for 
$\tau_X = T_L/2$ on the left column [Figs. \ref{Evst12}(a), (b), and (c)]
to the intracycle interference pattern in $\tau_X = T_L$ on the center column
[Figs. \ref{Evst12} (d), (e), and (f)] and the whole interference pattern for
$\tau_X = 2 T_L$ on the right column [Figs. \ref{Evst12} (g), (h), and (i)],
we corroborate the SCM prediction that the intrahalfcycle interference pattern
(spectrogram for $\tau_X = T_L/2$) works as a modulator of the intracycle pattern
(spectrogram for $\tau_X = T_L$), whereas, 
the latter does the same with the intercycle interference pattern or sidebands.


\section{\label{conc}Conclusions}


We have studied the electron emission produced by atomic hydrogen in its ground state
subject to an XUV pulse in the presence of an infrared laser pulse in the direction 
perpendicular to the common polarization axis of both pulses.
The previously developed SCM \cite{Gramajo16} for LAPE (XUV + IR)
in the forward direction has been extended for perpendicular emission.
In accordance to our recent study of LAPE in the forward direction \cite{Gramajo16},
the PE spectrum can be factorized as two contributions:
One accounting for sidebands formation and the other as a modulation.
Whereas the former can be interpreted as the intercycle interference of electron trajectories
from different optical cycles of the IR laser, 
the latter corresponds to intracycle interference stemming from the coherent superposition 
of four electron trajectories born in the same optical cycle.
Contrarily to parallel emission, the intracycle interference pattern for transversal emission
can be decomposed as the contribution of the two interfering trajectories
born within the same half optical period (\textit{intrahalfcycle} interference) 
and the Young-type interference between the contributions of the two half cycles into the same optical
cycle (\textit{interhalfcycle} interference).
We have shown that the electron trajectories born into the two half cycles
within the same optical cycle interfere destructively for the absorption
and/or emission of an even number of IR photons, which leads 
to the exchange of only an odd number of laser photons in the formation of the sidebands.
Therefore, the absorption line of the XUV photon alone (with no exchange of laser photons) 
is forbidden.
We show that the \textit{intrahalfcycle} interference pattern modulates the intracycle pattern,
which, in the same way, modulates the sidebands.
We have observed a very good agreement of our SCM energy spectrum with the corresponding one 
to the SFA and the \textit{ab initio} solution of the TDSE.

By studying the dependence of the electron emission on the laser intensity, we have observed that as the IR field increases the spectra becomes wider and approximately bounded within the classical energy domain. We can conclude that the SFA is accurate to describe the PE spectrum
perpendicular to the polarization direction, especially for low and moderate laser intensities so
that the electron ionization by the IR laser alone is low compared to LAPE.
Finally, by analyzing the electron spectrum as a function of the time delay between the two pulses $\Delta_X$, we have shown that the intrahalfcycle pattern is $\pi-$periodic in the optical phase with a
probability jump that reproduces the profile of the square of the laser vector potential.

\begin{acknowledgments}
Work supported by CONICET PIP0386, PICT-2012-3004 and PICT-2014-2363 of ANPCyT (Argentina), and the University of Buenos Aires (UBACyT 20020130100617BA).
\end{acknowledgments}

\bibliography{biblio}

\providecommand{\noopsort}[1]{}\providecommand{\singleletter}[1]{#1}%
\begin{thebibliography}{50}%
\makeatletter
\providecommand \@ifxundefined [1]{%
 \@ifx{#1\undefined}
}%
\providecommand \@ifnum [1]{%
 \ifnum #1\expandafter \@firstoftwo
 \else \expandafter \@secondoftwo
 \fi
}%
\providecommand \@ifx [1]{%
 \ifx #1\expandafter \@firstoftwo
 \else \expandafter \@secondoftwo
 \fi
}%
\providecommand \natexlab [1]{#1}%
\providecommand \enquote  [1]{``#1''}%
\providecommand \bibnamefont  [1]{#1}%
\providecommand \bibfnamefont [1]{#1}%
\providecommand \citenamefont [1]{#1}%
\providecommand \href@noop [0]{\@secondoftwo}%
\providecommand \href [0]{\begingroup \@sanitize@url \@href}%
\providecommand \@href[1]{\@@startlink{#1}\@@href}%
\providecommand \@@href[1]{\endgroup#1\@@endlink}%
\providecommand \@sanitize@url [0]{\catcode `\\12\catcode `\$12\catcode
  `\&12\catcode `\#12\catcode `\^12\catcode `\_12\catcode `\%12\relax}%
\providecommand \@@startlink[1]{}%
\providecommand \@@endlink[0]{}%
\providecommand \url  [0]{\begingroup\@sanitize@url \@url }%
\providecommand \@url [1]{\endgroup\@href {#1}{\urlprefix }}%
\providecommand \urlprefix  [0]{URL }%
\providecommand \Eprint [0]{\href }%
\providecommand \doibase [0]{http://dx.doi.org/}%
\providecommand \selectlanguage [0]{\@gobble}%
\providecommand \bibinfo  [0]{\@secondoftwo}%
\providecommand \bibfield  [0]{\@secondoftwo}%
\providecommand \translation [1]{[#1]}%
\providecommand \BibitemOpen [0]{}%
\providecommand \bibitemStop [0]{}%
\providecommand \bibitemNoStop [0]{.\EOS\space}%
\providecommand \EOS [0]{\spacefactor3000\relax}%
\providecommand \BibitemShut  [1]{\csname bibitem#1\endcsname}%
\let\auto@bib@innerbib\@empty
\bibitem [{\citenamefont {V\'eniard}\ \emph {et~al.}(1995)\citenamefont
  {V\'eniard}, \citenamefont {Ta\"{\i}eb},\ and\ \citenamefont
  {Maquet}}]{Veniard95}%
  \BibitemOpen
  \bibfield  {author} {\bibinfo {author} {\bibfnamefont {V.}~\bibnamefont
  {V\'eniard}}, \bibinfo {author} {\bibfnamefont {R.}~\bibnamefont
  {Ta\"{\i}eb}}, \ and\ \bibinfo {author} {\bibfnamefont {A.}~\bibnamefont
  {Maquet}},\ }\href {\doibase 10.1103/PhysRevLett.74.4161} {\bibfield
  {journal} {\bibinfo  {journal} {Phys. Rev. Lett.}\ }\textbf {\bibinfo
  {volume} {74}},\ \bibinfo {pages} {4161} (\bibinfo {year}
  {1995})}\BibitemShut {NoStop}%
\bibitem [{\citenamefont {Glover}\ \emph {et~al.}(1996)\citenamefont {Glover},
  \citenamefont {Schoenlein}, \citenamefont {Chin},\ and\ \citenamefont
  {Shank}}]{Glover1996}%
  \BibitemOpen
  \bibfield  {author} {\bibinfo {author} {\bibfnamefont {T.~E.}\ \bibnamefont
  {Glover}}, \bibinfo {author} {\bibfnamefont {R.~W.}\ \bibnamefont
  {Schoenlein}}, \bibinfo {author} {\bibfnamefont {A.~H.}\ \bibnamefont
  {Chin}}, \ and\ \bibinfo {author} {\bibfnamefont {C.~V.}\ \bibnamefont
  {Shank}},\ }\href {\doibase 10.1103/PhysRevLett.76.2468} {\bibfield
  {journal} {\bibinfo  {journal} {Phys. Rev. Lett.}\ }\textbf {\bibinfo
  {volume} {76}},\ \bibinfo {pages} {2468} (\bibinfo {year}
  {1996})}\BibitemShut {NoStop}%
\bibitem [{\citenamefont {Aseyev}\ \emph {et~al.}(2003)\citenamefont {Aseyev},
  \citenamefont {Ni}, \citenamefont {Frasinski}, \citenamefont {Muller},\ and\
  \citenamefont {Vrakking}}]{Aseyev2003}%
  \BibitemOpen
  \bibfield  {author} {\bibinfo {author} {\bibfnamefont {S.~A.}\ \bibnamefont
  {Aseyev}}, \bibinfo {author} {\bibfnamefont {Y.}~\bibnamefont {Ni}}, \bibinfo
  {author} {\bibfnamefont {L.~J.}\ \bibnamefont {Frasinski}}, \bibinfo {author}
  {\bibfnamefont {H.~G.}\ \bibnamefont {Muller}}, \ and\ \bibinfo {author}
  {\bibfnamefont {M.~J.~J.}\ \bibnamefont {Vrakking}},\ }\href {\doibase
  10.1103/PhysRevLett.91.223902} {\bibfield  {journal} {\bibinfo  {journal}
  {Phys. Rev. Lett.}\ }\textbf {\bibinfo {volume} {91}},\ \bibinfo {pages}
  {223902} (\bibinfo {year} {2003})}\BibitemShut {NoStop}%
\bibitem [{\citenamefont {O'Keeffe}\ \emph {et~al.}(2004)\citenamefont
  {O'Keeffe}, \citenamefont {L\'opez-Martens}, \citenamefont {Mauritsson},
  \citenamefont {Johansson}, \citenamefont {L'Huillier}, \citenamefont
  {V\'eniard}, \citenamefont {Ta\"{\i}eb}, \citenamefont {Maquet},\ and\
  \citenamefont {Meyer}}]{OKeeffe2004}%
  \BibitemOpen
  \bibfield  {author} {\bibinfo {author} {\bibfnamefont {P.}~\bibnamefont
  {O'Keeffe}}, \bibinfo {author} {\bibfnamefont {R.}~\bibnamefont
  {L\'opez-Martens}}, \bibinfo {author} {\bibfnamefont {J.}~\bibnamefont
  {Mauritsson}}, \bibinfo {author} {\bibfnamefont {A.}~\bibnamefont
  {Johansson}}, \bibinfo {author} {\bibfnamefont {A.}~\bibnamefont
  {L'Huillier}}, \bibinfo {author} {\bibfnamefont {V.}~\bibnamefont
  {V\'eniard}}, \bibinfo {author} {\bibfnamefont {R.}~\bibnamefont
  {Ta\"{\i}eb}}, \bibinfo {author} {\bibfnamefont {A.}~\bibnamefont {Maquet}},
  \ and\ \bibinfo {author} {\bibfnamefont {M.}~\bibnamefont {Meyer}},\ }\href
  {\doibase 10.1103/PhysRevA.69.051401} {\bibfield  {journal} {\bibinfo
  {journal} {Phys. Rev. A}\ }\textbf {\bibinfo {volume} {69}},\ \bibinfo
  {pages} {051401} (\bibinfo {year} {2004})}\BibitemShut {NoStop}%
\bibitem [{\citenamefont {Guy\'etand}\ \emph {et~al.}(2005)\citenamefont
  {Guy\'etand}, \citenamefont {Gisselbrecht}, \citenamefont {Huetz},
  \citenamefont {Agostini}, \citenamefont {Ta\"{\i}eb}, \citenamefont
  {V\'eniard}, \citenamefont {Maquet}, \citenamefont {Antonucci}, \citenamefont
  {Boyko}, \citenamefont {Valentin},\ and\ \citenamefont
  {Douillet}}]{Guyetand2005}%
  \BibitemOpen
  \bibfield  {author} {\bibinfo {author} {\bibfnamefont {O.}~\bibnamefont
  {Guy\'etand}}, \bibinfo {author} {\bibfnamefont {M.}~\bibnamefont
  {Gisselbrecht}}, \bibinfo {author} {\bibfnamefont {A.}~\bibnamefont {Huetz}},
  \bibinfo {author} {\bibfnamefont {P.}~\bibnamefont {Agostini}}, \bibinfo
  {author} {\bibfnamefont {R.}~\bibnamefont {Ta\"{\i}eb}}, \bibinfo {author}
  {\bibfnamefont {V.}~\bibnamefont {V\'eniard}}, \bibinfo {author}
  {\bibfnamefont {A.}~\bibnamefont {Maquet}}, \bibinfo {author} {\bibfnamefont
  {L.}~\bibnamefont {Antonucci}}, \bibinfo {author} {\bibfnamefont
  {O.}~\bibnamefont {Boyko}}, \bibinfo {author} {\bibfnamefont
  {C.}~\bibnamefont {Valentin}}, \ and\ \bibinfo {author} {\bibfnamefont
  {D.}~\bibnamefont {Douillet}},\ }\href
  {http://stacks.iop.org/0953-4075/38/i=22/a=L01} {\bibfield  {journal}
  {\bibinfo  {journal} {Journal of Physics B: Atomic, Molecular and Optical
  Physics}\ }\textbf {\bibinfo {volume} {38}},\ \bibinfo {pages} {L357}
  (\bibinfo {year} {2005})}\BibitemShut {NoStop}%
\bibitem [{\citenamefont {Meyer}\ \emph {et~al.}(2008)\citenamefont {Meyer},
  \citenamefont {Cubaynes}, \citenamefont {Glijer}, \citenamefont {Dardis},
  \citenamefont {Hayden}, \citenamefont {Hough}, \citenamefont {Richardson},
  \citenamefont {Kennedy}, \citenamefont {Costello}, \citenamefont {Radcliffe},
  \citenamefont {D\"usterer}, \citenamefont {Azima}, \citenamefont {Li},
  \citenamefont {Redlin}, \citenamefont {Feldhaus}, \citenamefont {Ta\"{\i}eb},
  \citenamefont {Maquet}, \citenamefont {Grum-Grzhimailo}, \citenamefont
  {Gryzlova},\ and\ \citenamefont {Strakhova}}]{Meyer2008}%
  \BibitemOpen
  \bibfield  {author} {\bibinfo {author} {\bibfnamefont {M.}~\bibnamefont
  {Meyer}}, \bibinfo {author} {\bibfnamefont {D.}~\bibnamefont {Cubaynes}},
  \bibinfo {author} {\bibfnamefont {D.}~\bibnamefont {Glijer}}, \bibinfo
  {author} {\bibfnamefont {J.}~\bibnamefont {Dardis}}, \bibinfo {author}
  {\bibfnamefont {P.}~\bibnamefont {Hayden}}, \bibinfo {author} {\bibfnamefont
  {P.}~\bibnamefont {Hough}}, \bibinfo {author} {\bibfnamefont
  {V.}~\bibnamefont {Richardson}}, \bibinfo {author} {\bibfnamefont {E.~T.}\
  \bibnamefont {Kennedy}}, \bibinfo {author} {\bibfnamefont {J.~T.}\
  \bibnamefont {Costello}}, \bibinfo {author} {\bibfnamefont {P.}~\bibnamefont
  {Radcliffe}}, \bibinfo {author} {\bibfnamefont {S.}~\bibnamefont
  {D\"usterer}}, \bibinfo {author} {\bibfnamefont {A.}~\bibnamefont {Azima}},
  \bibinfo {author} {\bibfnamefont {W.~B.}\ \bibnamefont {Li}}, \bibinfo
  {author} {\bibfnamefont {H.}~\bibnamefont {Redlin}}, \bibinfo {author}
  {\bibfnamefont {J.}~\bibnamefont {Feldhaus}}, \bibinfo {author}
  {\bibfnamefont {R.}~\bibnamefont {Ta\"{\i}eb}}, \bibinfo {author}
  {\bibfnamefont {A.}~\bibnamefont {Maquet}}, \bibinfo {author} {\bibfnamefont
  {A.~N.}\ \bibnamefont {Grum-Grzhimailo}}, \bibinfo {author} {\bibfnamefont
  {E.~V.}\ \bibnamefont {Gryzlova}}, \ and\ \bibinfo {author} {\bibfnamefont
  {S.~I.}\ \bibnamefont {Strakhova}},\ }\href {\doibase
  10.1103/PhysRevLett.101.193002} {\bibfield  {journal} {\bibinfo  {journal}
  {Phys. Rev. Lett.}\ }\textbf {\bibinfo {volume} {101}},\ \bibinfo {pages}
  {193002} (\bibinfo {year} {2008})}\BibitemShut {NoStop}%
\bibitem [{\citenamefont {Meyer}\ \emph {et~al.}(2010)\citenamefont {Meyer},
  \citenamefont {Costello}, \citenamefont {D\"usterer}, \citenamefont {Li},\
  and\ \citenamefont {Radcliffe}}]{Meyer2010JPB}%
  \BibitemOpen
  \bibfield  {author} {\bibinfo {author} {\bibfnamefont {M.}~\bibnamefont
  {Meyer}}, \bibinfo {author} {\bibfnamefont {J.~T.}\ \bibnamefont {Costello}},
  \bibinfo {author} {\bibfnamefont {S.}~\bibnamefont {D\"usterer}}, \bibinfo
  {author} {\bibfnamefont {W.~B.}\ \bibnamefont {Li}}, \ and\ \bibinfo {author}
  {\bibfnamefont {P.}~\bibnamefont {Radcliffe}},\ }\href
  {http://stacks.iop.org/0953-4075/43/i=19/a=194006} {\bibfield  {journal}
  {\bibinfo  {journal} {Journal of Physics B: Atomic, Molecular and Optical
  Physics}\ }\textbf {\bibinfo {volume} {43}},\ \bibinfo {pages} {194006}
  (\bibinfo {year} {2010})}\BibitemShut {NoStop}%
\bibitem [{\citenamefont {Radcliffe}\ \emph {et~al.}(2012)\citenamefont
  {Radcliffe}, \citenamefont {Arbeiter}, \citenamefont {Li}, \citenamefont
  {D\"usterer}, \citenamefont {Redlin}, \citenamefont {Hayden}, \citenamefont
  {Hough}, \citenamefont {Richardson}, \citenamefont {Costello}, \citenamefont
  {Fennel},\ and\ \citenamefont {Meyer}}]{Radcliffe2012}%
  \BibitemOpen
  \bibfield  {author} {\bibinfo {author} {\bibfnamefont {P.}~\bibnamefont
  {Radcliffe}}, \bibinfo {author} {\bibfnamefont {M.}~\bibnamefont {Arbeiter}},
  \bibinfo {author} {\bibfnamefont {W.~B.}\ \bibnamefont {Li}}, \bibinfo
  {author} {\bibfnamefont {S.}~\bibnamefont {D\"usterer}}, \bibinfo {author}
  {\bibfnamefont {H.}~\bibnamefont {Redlin}}, \bibinfo {author} {\bibfnamefont
  {P.}~\bibnamefont {Hayden}}, \bibinfo {author} {\bibfnamefont
  {P.}~\bibnamefont {Hough}}, \bibinfo {author} {\bibfnamefont
  {V.}~\bibnamefont {Richardson}}, \bibinfo {author} {\bibfnamefont {J.~T.}\
  \bibnamefont {Costello}}, \bibinfo {author} {\bibfnamefont {T.}~\bibnamefont
  {Fennel}}, \ and\ \bibinfo {author} {\bibfnamefont {M.}~\bibnamefont
  {Meyer}},\ }\href {http://stacks.iop.org/1367-2630/14/i=4/a=043008}
  {\bibfield  {journal} {\bibinfo  {journal} {New Journal of Physics}\ }\textbf
  {\bibinfo {volume} {14}},\ \bibinfo {pages} {043008} (\bibinfo {year}
  {2012})}\BibitemShut {NoStop}%
\bibitem [{\citenamefont {Mazza}\ \emph {et~al.}(2014)\citenamefont {Mazza},
  \citenamefont {Ilchen}, \citenamefont {Rafipoor}, \citenamefont {Callegari},
  \citenamefont {Finetti}, \citenamefont {Plekan}, \citenamefont {Prince},
  \citenamefont {Richter}, \citenamefont {Danailov}, \citenamefont
  {Demidovich}, \citenamefont {De~Ninno}, \citenamefont {Grazioli},
  \citenamefont {Ivanov}, \citenamefont {Mahne}, \citenamefont {Raimondi},
  \citenamefont {Svetina}, \citenamefont {Avaldi}, \citenamefont {Bolognesi},
  \citenamefont {Coreno}, \citenamefont {O'Keeffe}, \citenamefont {Di~Fraia},
  \citenamefont {Devetta}, \citenamefont {Ovcharenko}, \citenamefont
  {M\"oller}, \citenamefont {Lyamayev}, \citenamefont {Stienkemeier},
  \citenamefont {D\"usterer}, \citenamefont {Ueda}, \citenamefont {Costello},
  \citenamefont {Kazansky}, \citenamefont {Kabachnik},\ and\ \citenamefont
  {Meyer}}]{Mazza2014}%
  \BibitemOpen
  \bibfield  {author} {\bibinfo {author} {\bibfnamefont {T.}~\bibnamefont
  {Mazza}}, \bibinfo {author} {\bibfnamefont {M.}~\bibnamefont {Ilchen}},
  \bibinfo {author} {\bibfnamefont {A.~J.}\ \bibnamefont {Rafipoor}}, \bibinfo
  {author} {\bibfnamefont {C.}~\bibnamefont {Callegari}}, \bibinfo {author}
  {\bibfnamefont {P.}~\bibnamefont {Finetti}}, \bibinfo {author} {\bibfnamefont
  {O.}~\bibnamefont {Plekan}}, \bibinfo {author} {\bibfnamefont {K.~C.}\
  \bibnamefont {Prince}}, \bibinfo {author} {\bibfnamefont {R.}~\bibnamefont
  {Richter}}, \bibinfo {author} {\bibfnamefont {M.~B.}\ \bibnamefont
  {Danailov}}, \bibinfo {author} {\bibfnamefont {A.}~\bibnamefont
  {Demidovich}}, \bibinfo {author} {\bibfnamefont {G.}~\bibnamefont
  {De~Ninno}}, \bibinfo {author} {\bibfnamefont {C.}~\bibnamefont {Grazioli}},
  \bibinfo {author} {\bibfnamefont {R.}~\bibnamefont {Ivanov}}, \bibinfo
  {author} {\bibfnamefont {N.}~\bibnamefont {Mahne}}, \bibinfo {author}
  {\bibfnamefont {L.}~\bibnamefont {Raimondi}}, \bibinfo {author}
  {\bibfnamefont {C.}~\bibnamefont {Svetina}}, \bibinfo {author} {\bibfnamefont
  {L.}~\bibnamefont {Avaldi}}, \bibinfo {author} {\bibfnamefont
  {P.}~\bibnamefont {Bolognesi}}, \bibinfo {author} {\bibfnamefont
  {M.}~\bibnamefont {Coreno}}, \bibinfo {author} {\bibfnamefont
  {P.}~\bibnamefont {O'Keeffe}}, \bibinfo {author} {\bibfnamefont
  {M.}~\bibnamefont {Di~Fraia}}, \bibinfo {author} {\bibfnamefont
  {M.}~\bibnamefont {Devetta}}, \bibinfo {author} {\bibfnamefont
  {Y.}~\bibnamefont {Ovcharenko}}, \bibinfo {author} {\bibfnamefont
  {T.}~\bibnamefont {M\"oller}}, \bibinfo {author} {\bibfnamefont
  {V.}~\bibnamefont {Lyamayev}}, \bibinfo {author} {\bibfnamefont
  {F.}~\bibnamefont {Stienkemeier}}, \bibinfo {author} {\bibfnamefont
  {S.}~\bibnamefont {D\"usterer}}, \bibinfo {author} {\bibfnamefont
  {K.}~\bibnamefont {Ueda}}, \bibinfo {author} {\bibfnamefont {J.~T.}\
  \bibnamefont {Costello}}, \bibinfo {author} {\bibfnamefont {A.~K.}\
  \bibnamefont {Kazansky}}, \bibinfo {author} {\bibfnamefont {N.~M.}\
  \bibnamefont {Kabachnik}}, \ and\ \bibinfo {author} {\bibfnamefont
  {M.}~\bibnamefont {Meyer}},\ }\href {http://dx.doi.org/10.1038/ncomms4648}
  {\bibfield  {journal} {\bibinfo  {journal} {Nature Communications}\ }\textbf
  {\bibinfo {volume} {5}},\ \bibinfo {pages} {3648} (\bibinfo {year} {2014})},\
  \Eprint {http://arxiv.org/abs/http://dx.doi.org/10.1038/ncomms4648}
  {http://dx.doi.org/10.1038/ncomms4648} \BibitemShut {NoStop}%
\bibitem [{\citenamefont {Hayden}\ \emph {et~al.}(2016)\citenamefont {Hayden},
  \citenamefont {Dardis}, \citenamefont {Hough}, \citenamefont {Richardson},
  \citenamefont {Kennedy}, \citenamefont {Costello}, \citenamefont
  {D\"usterer}, \citenamefont {Redlin}, \citenamefont {Feldhaus}, \citenamefont
  {Li}, \citenamefont {Cubaynes},\ and\ \citenamefont {Meyer}}]{Hayden16}%
  \BibitemOpen
  \bibfield  {author} {\bibinfo {author} {\bibfnamefont {P.}~\bibnamefont
  {Hayden}}, \bibinfo {author} {\bibfnamefont {J.}~\bibnamefont {Dardis}},
  \bibinfo {author} {\bibfnamefont {P.}~\bibnamefont {Hough}}, \bibinfo
  {author} {\bibfnamefont {V.}~\bibnamefont {Richardson}}, \bibinfo {author}
  {\bibfnamefont {E.~T.}\ \bibnamefont {Kennedy}}, \bibinfo {author}
  {\bibfnamefont {J.~T.}\ \bibnamefont {Costello}}, \bibinfo {author}
  {\bibfnamefont {S.}~\bibnamefont {D\"usterer}}, \bibinfo {author}
  {\bibfnamefont {H.}~\bibnamefont {Redlin}}, \bibinfo {author} {\bibfnamefont
  {J.}~\bibnamefont {Feldhaus}}, \bibinfo {author} {\bibfnamefont {W.~B.}\
  \bibnamefont {Li}}, \bibinfo {author} {\bibfnamefont {D.}~\bibnamefont
  {Cubaynes}}, \ and\ \bibinfo {author} {\bibfnamefont {M.}~\bibnamefont
  {Meyer}},\ }\href {\doibase 10.1080/09500340.2015.1117669} {\bibfield
  {journal} {\bibinfo  {journal} {Journal of Modern Optics}\ }\textbf {\bibinfo
  {volume} {63}},\ \bibinfo {pages} {358} (\bibinfo {year} {2016})},\ \Eprint
  {http://arxiv.org/abs/http://dx.doi.org/10.1080/09500340.2015.1117669}
  {http://dx.doi.org/10.1080/09500340.2015.1117669} \BibitemShut {NoStop}%
\bibitem [{\citenamefont {D\"usterer}\ \emph {et~al.}(2016)\citenamefont
  {D\"usterer}, \citenamefont {Hartmann}, \citenamefont {Babies}, \citenamefont
  {Beckmann}, \citenamefont {Brenner}, \citenamefont {Buck}, \citenamefont
  {Costello}, \citenamefont {Dammann}, \citenamefont {Fanis}, \citenamefont
  {Ge{\ss}ler}, \citenamefont {Glaser}, \citenamefont {Ilchen}, \citenamefont
  {Johnsson}, \citenamefont {Kazansky}, \citenamefont {Kelly}, \citenamefont
  {Mazza}, \citenamefont {Meyer}, \citenamefont {Nosik}, \citenamefont
  {Sazhina}, \citenamefont {Scholz}, \citenamefont {Seltmann}, \citenamefont
  {Sotoudi}, \citenamefont {Viefhaus},\ and\ \citenamefont
  {Kabachnik}}]{Dusterer2016}%
  \BibitemOpen
  \bibfield  {author} {\bibinfo {author} {\bibfnamefont {S.}~\bibnamefont
  {D\"usterer}}, \bibinfo {author} {\bibfnamefont {G.}~\bibnamefont
  {Hartmann}}, \bibinfo {author} {\bibfnamefont {F.}~\bibnamefont {Babies}},
  \bibinfo {author} {\bibfnamefont {A.}~\bibnamefont {Beckmann}}, \bibinfo
  {author} {\bibfnamefont {G.}~\bibnamefont {Brenner}}, \bibinfo {author}
  {\bibfnamefont {J.}~\bibnamefont {Buck}}, \bibinfo {author} {\bibfnamefont
  {J.}~\bibnamefont {Costello}}, \bibinfo {author} {\bibfnamefont
  {L.}~\bibnamefont {Dammann}}, \bibinfo {author} {\bibfnamefont {A.~D.}\
  \bibnamefont {Fanis}}, \bibinfo {author} {\bibfnamefont {P.}~\bibnamefont
  {Ge{\ss}ler}}, \bibinfo {author} {\bibfnamefont {L.}~\bibnamefont {Glaser}},
  \bibinfo {author} {\bibfnamefont {M.}~\bibnamefont {Ilchen}}, \bibinfo
  {author} {\bibfnamefont {P.}~\bibnamefont {Johnsson}}, \bibinfo {author}
  {\bibfnamefont {A.~K.}\ \bibnamefont {Kazansky}}, \bibinfo {author}
  {\bibfnamefont {T.~J.}\ \bibnamefont {Kelly}}, \bibinfo {author}
  {\bibfnamefont {T.}~\bibnamefont {Mazza}}, \bibinfo {author} {\bibfnamefont
  {M.}~\bibnamefont {Meyer}}, \bibinfo {author} {\bibfnamefont {V.~L.}\
  \bibnamefont {Nosik}}, \bibinfo {author} {\bibfnamefont {I.~P.}\ \bibnamefont
  {Sazhina}}, \bibinfo {author} {\bibfnamefont {F.}~\bibnamefont {Scholz}},
  \bibinfo {author} {\bibfnamefont {J.}~\bibnamefont {Seltmann}}, \bibinfo
  {author} {\bibfnamefont {H.}~\bibnamefont {Sotoudi}}, \bibinfo {author}
  {\bibfnamefont {J.}~\bibnamefont {Viefhaus}}, \ and\ \bibinfo {author}
  {\bibfnamefont {N.~M.}\ \bibnamefont {Kabachnik}},\ }\href
  {http://stacks.iop.org/0953-4075/49/i=16/a=165003} {\bibfield  {journal}
  {\bibinfo  {journal} {Journal of Physics B: Atomic, Molecular and Optical
  Physics}\ }\textbf {\bibinfo {volume} {49}},\ \bibinfo {pages} {165003}
  (\bibinfo {year} {2016})}\BibitemShut {NoStop}%
\bibitem [{\citenamefont {Kienberger}\ \emph {et~al.}(2002)\citenamefont
  {Kienberger}, \citenamefont {Hentschel}, \citenamefont {Uiberacker},
  \citenamefont {Spielmann}, \citenamefont {Kitzler}, \citenamefont {Scrinzi},
  \citenamefont {Wieland}, \citenamefont {Westerwalbesloh}, \citenamefont
  {Kleineberg}, \citenamefont {Heinzmann}, \citenamefont {Drescher},\ and\
  \citenamefont {Krausz}}]{Kienberger2002}%
  \BibitemOpen
  \bibfield  {author} {\bibinfo {author} {\bibfnamefont {R.}~\bibnamefont
  {Kienberger}}, \bibinfo {author} {\bibfnamefont {M.}~\bibnamefont
  {Hentschel}}, \bibinfo {author} {\bibfnamefont {M.}~\bibnamefont
  {Uiberacker}}, \bibinfo {author} {\bibfnamefont {C.}~\bibnamefont
  {Spielmann}}, \bibinfo {author} {\bibfnamefont {M.}~\bibnamefont {Kitzler}},
  \bibinfo {author} {\bibfnamefont {A.}~\bibnamefont {Scrinzi}}, \bibinfo
  {author} {\bibfnamefont {M.}~\bibnamefont {Wieland}}, \bibinfo {author}
  {\bibfnamefont {T.}~\bibnamefont {Westerwalbesloh}}, \bibinfo {author}
  {\bibfnamefont {U.}~\bibnamefont {Kleineberg}}, \bibinfo {author}
  {\bibfnamefont {U.}~\bibnamefont {Heinzmann}}, \bibinfo {author}
  {\bibfnamefont {M.}~\bibnamefont {Drescher}}, \ and\ \bibinfo {author}
  {\bibfnamefont {F.}~\bibnamefont {Krausz}},\ }\href {\doibase
  10.1126/science.1073866} {\bibfield  {journal} {\bibinfo  {journal}
  {Science}\ }\textbf {\bibinfo {volume} {297}},\ \bibinfo {pages} {1144}
  (\bibinfo {year} {2002})},\ \Eprint
  {http://arxiv.org/abs/http://science.sciencemag.org/content/297/5584/1144.full.pdf}
  {http://science.sciencemag.org/content/297/5584/1144.full.pdf} \BibitemShut
  {NoStop}%
\bibitem [{\citenamefont {{Drescher}}\ and\ \citenamefont
  {{Krausz}}(2005)}]{Drescher05}%
  \BibitemOpen
  \bibfield  {author} {\bibinfo {author} {\bibfnamefont {M.}~\bibnamefont
  {{Drescher}}}\ and\ \bibinfo {author} {\bibfnamefont {F.}~\bibnamefont
  {{Krausz}}},\ }\href {\doibase 10.1088/0953-4075/38/9/019} {\bibfield
  {journal} {\bibinfo  {journal} {Journal of Physics B Atomic Molecular
  Physics}\ }\textbf {\bibinfo {volume} {38}},\ \bibinfo {pages} {S727}
  (\bibinfo {year} {2005})}\BibitemShut {NoStop}%
\bibitem [{\citenamefont {Maquet}\ and\ \citenamefont
  {Ta\"{\i}eb}(2007)}]{Maquet2007}%
  \BibitemOpen
  \bibfield  {author} {\bibinfo {author} {\bibfnamefont {A.}~\bibnamefont
  {Maquet}}\ and\ \bibinfo {author} {\bibfnamefont {R.}~\bibnamefont
  {Ta\"{\i}eb}},\ }\href {\doibase 10.1080/09500340701306751} {\bibfield
  {journal} {\bibinfo  {journal} {Journal of Modern Optics}\ }\textbf {\bibinfo
  {volume} {54}},\ \bibinfo {pages} {1847} (\bibinfo {year} {2007})},\ \Eprint
  {http://arxiv.org/abs/http://dx.doi.org/10.1080/09500340701306751}
  {http://dx.doi.org/10.1080/09500340701306751} \BibitemShut {NoStop}%
\bibitem [{\citenamefont {D\"usterer}\ \emph {et~al.}(2013)\citenamefont
  {D\"usterer}, \citenamefont {Rading}, \citenamefont {Johnsson}, \citenamefont
  {Rouz\'ee}, \citenamefont {Hundertmark}, \citenamefont {Vrakking},
  \citenamefont {Radcliffe}, \citenamefont {Meyer}, \citenamefont {Kazansky},\
  and\ \citenamefont {Kabachnik}}]{Dusterer2013}%
  \BibitemOpen
  \bibfield  {author} {\bibinfo {author} {\bibfnamefont {S.}~\bibnamefont
  {D\"usterer}}, \bibinfo {author} {\bibfnamefont {L.}~\bibnamefont {Rading}},
  \bibinfo {author} {\bibfnamefont {P.}~\bibnamefont {Johnsson}}, \bibinfo
  {author} {\bibfnamefont {A.}~\bibnamefont {Rouz\'ee}}, \bibinfo {author}
  {\bibfnamefont {A.}~\bibnamefont {Hundertmark}}, \bibinfo {author}
  {\bibfnamefont {M.~J.~J.}\ \bibnamefont {Vrakking}}, \bibinfo {author}
  {\bibfnamefont {P.}~\bibnamefont {Radcliffe}}, \bibinfo {author}
  {\bibfnamefont {M.}~\bibnamefont {Meyer}}, \bibinfo {author} {\bibfnamefont
  {A.~K.}\ \bibnamefont {Kazansky}}, \ and\ \bibinfo {author} {\bibfnamefont
  {N.~M.}\ \bibnamefont {Kabachnik}},\ }\href
  {http://stacks.iop.org/0953-4075/46/i=16/a=164026} {\bibfield  {journal}
  {\bibinfo  {journal} {Journal of Physics B: Atomic, Molecular and Optical
  Physics}\ }\textbf {\bibinfo {volume} {46}},\ \bibinfo {pages} {164026}
  (\bibinfo {year} {2013})}\BibitemShut {NoStop}%
\bibitem [{\citenamefont {Itatani}\ \emph {et~al.}(2002)\citenamefont
  {Itatani}, \citenamefont {Qu\'er\'e}, \citenamefont {Yudin}, \citenamefont
  {Ivanov}, \citenamefont {Krausz},\ and\ \citenamefont {Corkum}}]{Itatani02}%
  \BibitemOpen
  \bibfield  {author} {\bibinfo {author} {\bibfnamefont {J.}~\bibnamefont
  {Itatani}}, \bibinfo {author} {\bibfnamefont {F.}~\bibnamefont {Qu\'er\'e}},
  \bibinfo {author} {\bibfnamefont {G.~L.}\ \bibnamefont {Yudin}}, \bibinfo
  {author} {\bibfnamefont {M.~Y.}\ \bibnamefont {Ivanov}}, \bibinfo {author}
  {\bibfnamefont {F.}~\bibnamefont {Krausz}}, \ and\ \bibinfo {author}
  {\bibfnamefont {P.~B.}\ \bibnamefont {Corkum}},\ }\href {\doibase
  10.1103/PhysRevLett.88.173903} {\bibfield  {journal} {\bibinfo  {journal}
  {Phys. Rev. Lett.}\ }\textbf {\bibinfo {volume} {88}},\ \bibinfo {pages}
  {173903} (\bibinfo {year} {2002})}\BibitemShut {NoStop}%
\bibitem [{\citenamefont {Fr\"uhling}\ \emph {et~al.}(2009)\citenamefont
  {Fr\"uhling}, \citenamefont {Wieland}, \citenamefont {Gensch}, \citenamefont
  {Gebert}, \citenamefont {Schutte}, \citenamefont {Krikunova}, \citenamefont
  {Kalms}, \citenamefont {Budzyn}, \citenamefont {Grimm}, \citenamefont
  {Rossbach}, \citenamefont {Plonjes},\ and\ \citenamefont
  {Drescher}}]{Fruehling09}%
  \BibitemOpen
  \bibfield  {author} {\bibinfo {author} {\bibfnamefont {U.}~\bibnamefont
  {Fr\"uhling}}, \bibinfo {author} {\bibfnamefont {M.}~\bibnamefont {Wieland}},
  \bibinfo {author} {\bibfnamefont {M.}~\bibnamefont {Gensch}}, \bibinfo
  {author} {\bibfnamefont {T.}~\bibnamefont {Gebert}}, \bibinfo {author}
  {\bibfnamefont {B.}~\bibnamefont {Schutte}}, \bibinfo {author} {\bibfnamefont
  {M.}~\bibnamefont {Krikunova}}, \bibinfo {author} {\bibfnamefont
  {R.}~\bibnamefont {Kalms}}, \bibinfo {author} {\bibfnamefont
  {F.}~\bibnamefont {Budzyn}}, \bibinfo {author} {\bibfnamefont
  {O.}~\bibnamefont {Grimm}}, \bibinfo {author} {\bibfnamefont
  {J.}~\bibnamefont {Rossbach}}, \bibinfo {author} {\bibfnamefont
  {E.}~\bibnamefont {Plonjes}}, \ and\ \bibinfo {author} {\bibfnamefont
  {M.}~\bibnamefont {Drescher}},\ }\href {\doibase 10.1038/nphoton.2009.160}
  {\bibfield  {journal} {\bibinfo  {journal} {Nat Photon}\ }\textbf {\bibinfo
  {volume} {3}},\ \bibinfo {pages} {523} (\bibinfo {year} {2009})}\BibitemShut
  {NoStop}%
\bibitem [{\citenamefont {Krausz}\ and\ \citenamefont
  {Ivanov}(2009)}]{Krausz2009}%
  \BibitemOpen
  \bibfield  {author} {\bibinfo {author} {\bibfnamefont {F.}~\bibnamefont
  {Krausz}}\ and\ \bibinfo {author} {\bibfnamefont {M.}~\bibnamefont
  {Ivanov}},\ }\href {\doibase 10.1103/RevModPhys.81.163} {\bibfield  {journal}
  {\bibinfo  {journal} {Rev. Mod. Phys.}\ }\textbf {\bibinfo {volume} {81}},\
  \bibinfo {pages} {163} (\bibinfo {year} {2009})}\BibitemShut {NoStop}%
\bibitem [{\citenamefont {{Pazourek}}\ \emph {et~al.}(2015)\citenamefont
  {{Pazourek}}, \citenamefont {{Nagele}},\ and\ \citenamefont
  {{Burgd{\"o}rfer}}}]{Pazourek15}%
  \BibitemOpen
  \bibfield  {author} {\bibinfo {author} {\bibfnamefont {R.}~\bibnamefont
  {{Pazourek}}}, \bibinfo {author} {\bibfnamefont {S.}~\bibnamefont
  {{Nagele}}}, \ and\ \bibinfo {author} {\bibfnamefont {J.}~\bibnamefont
  {{Burgd{\"o}rfer}}},\ }\href {\doibase 10.1103/RevModPhys.87.765} {\bibfield
  {journal} {\bibinfo  {journal} {Reviews of Modern Physics}\ }\textbf
  {\bibinfo {volume} {87}},\ \bibinfo {pages} {765} (\bibinfo {year}
  {2015})}\BibitemShut {NoStop}%
\bibitem [{\citenamefont {Ta\"{\i}eb}\ \emph {et~al.}(2000)\citenamefont
  {Ta\"{\i}eb}, \citenamefont {V\'eniard}, \citenamefont {Maquet},
  \citenamefont {Manakov},\ and\ \citenamefont {Marmo}}]{Taieb2000}%
  \BibitemOpen
  \bibfield  {author} {\bibinfo {author} {\bibfnamefont {R.}~\bibnamefont
  {Ta\"{\i}eb}}, \bibinfo {author} {\bibfnamefont {V.}~\bibnamefont
  {V\'eniard}}, \bibinfo {author} {\bibfnamefont {A.}~\bibnamefont {Maquet}},
  \bibinfo {author} {\bibfnamefont {N.~L.}\ \bibnamefont {Manakov}}, \ and\
  \bibinfo {author} {\bibfnamefont {S.~I.}\ \bibnamefont {Marmo}},\ }\href
  {\doibase 10.1103/PhysRevA.62.013402} {\bibfield  {journal} {\bibinfo
  {journal} {Phys. Rev. A}\ }\textbf {\bibinfo {volume} {62}},\ \bibinfo
  {pages} {013402} (\bibinfo {year} {2000})}\BibitemShut {NoStop}%
\bibitem [{\citenamefont {Kazansky}\ \emph {et~al.}(2012)\citenamefont
  {Kazansky}, \citenamefont {Grigorieva},\ and\ \citenamefont
  {Kabachnik}}]{Kazansky2012}%
  \BibitemOpen
  \bibfield  {author} {\bibinfo {author} {\bibfnamefont {A.~K.}\ \bibnamefont
  {Kazansky}}, \bibinfo {author} {\bibfnamefont {A.~V.}\ \bibnamefont
  {Grigorieva}}, \ and\ \bibinfo {author} {\bibfnamefont {N.~M.}\ \bibnamefont
  {Kabachnik}},\ }\href {\doibase 10.1103/PhysRevA.85.053409} {\bibfield
  {journal} {\bibinfo  {journal} {Phys. Rev. A}\ }\textbf {\bibinfo {volume}
  {85}},\ \bibinfo {pages} {053409} (\bibinfo {year} {2012})}\BibitemShut
  {NoStop}%
\bibitem [{\citenamefont {Kazansky}\ \emph {et~al.}(2014)\citenamefont
  {Kazansky}, \citenamefont {Bozhevolnov}, \citenamefont {Sazhina},\ and\
  \citenamefont {Kabachnik}}]{Kazansky2014}%
  \BibitemOpen
  \bibfield  {author} {\bibinfo {author} {\bibfnamefont {A.~K.}\ \bibnamefont
  {Kazansky}}, \bibinfo {author} {\bibfnamefont {A.~V.}\ \bibnamefont
  {Bozhevolnov}}, \bibinfo {author} {\bibfnamefont {I.~P.}\ \bibnamefont
  {Sazhina}}, \ and\ \bibinfo {author} {\bibfnamefont {N.~M.}\ \bibnamefont
  {Kabachnik}},\ }\href {http://stacks.iop.org/0953-4075/47/i=6/a=065602}
  {\bibfield  {journal} {\bibinfo  {journal} {Journal of Physics B: Atomic,
  Molecular and Optical Physics}\ }\textbf {\bibinfo {volume} {47}},\ \bibinfo
  {pages} {065602} (\bibinfo {year} {2014})}\BibitemShut {NoStop}%
\bibitem [{\citenamefont {Mazza}\ \emph {et~al.}(2015)\citenamefont {Mazza},
  \citenamefont {Gryzlova}, \citenamefont {Grum-Grzhimailo}, \citenamefont
  {Kazansky}, \citenamefont {Kabachnik},\ and\ \citenamefont
  {Meyer}}]{Mazza15}%
  \BibitemOpen
  \bibfield  {author} {\bibinfo {author} {\bibfnamefont {T.}~\bibnamefont
  {Mazza}}, \bibinfo {author} {\bibfnamefont {E.}~\bibnamefont {Gryzlova}},
  \bibinfo {author} {\bibfnamefont {A.}~\bibnamefont {Grum-Grzhimailo}},
  \bibinfo {author} {\bibfnamefont {A.}~\bibnamefont {Kazansky}}, \bibinfo
  {author} {\bibfnamefont {N.}~\bibnamefont {Kabachnik}}, \ and\ \bibinfo
  {author} {\bibfnamefont {M.}~\bibnamefont {Meyer}},\ }\href {\doibase
  http://dx.doi.org/10.1016/j.elspec.2015.08.011} {\bibfield  {journal}
  {\bibinfo  {journal} {Journal of Electron Spectroscopy and Related
  Phenomena}\ }\textbf {\bibinfo {volume} {204, Part B}},\ \bibinfo {pages}
  {313 } (\bibinfo {year} {2015})}\BibitemShut {NoStop}%
\bibitem [{\citenamefont {Richter}\ \emph {et~al.}(2015)\citenamefont
  {Richter}, \citenamefont {Kunitski}, \citenamefont {Sch\"offler},
  \citenamefont {Jahnke}, \citenamefont {Schmidt}, \citenamefont {Li},
  \citenamefont {Liu},\ and\ \citenamefont {D\"orner}}]{Richter2015}%
  \BibitemOpen
  \bibfield  {author} {\bibinfo {author} {\bibfnamefont {M.}~\bibnamefont
  {Richter}}, \bibinfo {author} {\bibfnamefont {M.}~\bibnamefont {Kunitski}},
  \bibinfo {author} {\bibfnamefont {M.}~\bibnamefont {Sch\"offler}}, \bibinfo
  {author} {\bibfnamefont {T.}~\bibnamefont {Jahnke}}, \bibinfo {author}
  {\bibfnamefont {L.~P.~H.}\ \bibnamefont {Schmidt}}, \bibinfo {author}
  {\bibfnamefont {M.}~\bibnamefont {Li}}, \bibinfo {author} {\bibfnamefont
  {Y.}~\bibnamefont {Liu}}, \ and\ \bibinfo {author} {\bibfnamefont
  {R.}~\bibnamefont {D\"orner}},\ }\href {\doibase
  10.1103/PhysRevLett.114.143001} {\bibfield  {journal} {\bibinfo  {journal}
  {Phys. Rev. Lett.}\ }\textbf {\bibinfo {volume} {114}},\ \bibinfo {pages}
  {143001} (\bibinfo {year} {2015})}\BibitemShut {NoStop}%
\bibitem [{\citenamefont {Haber}\ \emph {et~al.}(2009)\citenamefont {Haber},
  \citenamefont {Doughty},\ and\ \citenamefont {Leone}}]{Haber2009}%
  \BibitemOpen
  \bibfield  {author} {\bibinfo {author} {\bibfnamefont {L.~H.}\ \bibnamefont
  {Haber}}, \bibinfo {author} {\bibfnamefont {B.}~\bibnamefont {Doughty}}, \
  and\ \bibinfo {author} {\bibfnamefont {S.~R.}\ \bibnamefont {Leone}},\ }\href
  {\doibase 10.1021/jp903231n} {\bibfield  {journal} {\bibinfo  {journal} {The
  Journal of Physical Chemistry A}\ }\textbf {\bibinfo {volume} {113}},\
  \bibinfo {pages} {13152} (\bibinfo {year} {2009})},\ \bibinfo {note} {pMID:
  19610629},\ \Eprint
  {http://arxiv.org/abs/http://dx.doi.org/10.1021/jp903231n}
  {http://dx.doi.org/10.1021/jp903231n} \BibitemShut {NoStop}%
\bibitem [{\citenamefont {Mondal}\ \emph {et~al.}(2014)\citenamefont {Mondal},
  \citenamefont {Fukuzawa}, \citenamefont {Motomura}, \citenamefont
  {Tachibana}, \citenamefont {Nagaya}, \citenamefont {Sakai}, \citenamefont
  {Matsunami}, \citenamefont {Yase}, \citenamefont {Yao}, \citenamefont {Wada},
  \citenamefont {Hayashita}, \citenamefont {Saito}, \citenamefont {Callegari},
  \citenamefont {Prince}, \citenamefont {Miron}, \citenamefont {Nagasono},
  \citenamefont {Togashi}, \citenamefont {Yabashi}, \citenamefont {Ishikawa},
  \citenamefont {Kazansky}, \citenamefont {Kabachnik},\ and\ \citenamefont
  {Ueda}}]{Mondal2014}%
  \BibitemOpen
  \bibfield  {author} {\bibinfo {author} {\bibfnamefont {S.}~\bibnamefont
  {Mondal}}, \bibinfo {author} {\bibfnamefont {H.}~\bibnamefont {Fukuzawa}},
  \bibinfo {author} {\bibfnamefont {K.}~\bibnamefont {Motomura}}, \bibinfo
  {author} {\bibfnamefont {T.}~\bibnamefont {Tachibana}}, \bibinfo {author}
  {\bibfnamefont {K.}~\bibnamefont {Nagaya}}, \bibinfo {author} {\bibfnamefont
  {T.}~\bibnamefont {Sakai}}, \bibinfo {author} {\bibfnamefont
  {K.}~\bibnamefont {Matsunami}}, \bibinfo {author} {\bibfnamefont
  {S.}~\bibnamefont {Yase}}, \bibinfo {author} {\bibfnamefont {M.}~\bibnamefont
  {Yao}}, \bibinfo {author} {\bibfnamefont {S.}~\bibnamefont {Wada}}, \bibinfo
  {author} {\bibfnamefont {H.}~\bibnamefont {Hayashita}}, \bibinfo {author}
  {\bibfnamefont {N.}~\bibnamefont {Saito}}, \bibinfo {author} {\bibfnamefont
  {C.}~\bibnamefont {Callegari}}, \bibinfo {author} {\bibfnamefont {K.~C.}\
  \bibnamefont {Prince}}, \bibinfo {author} {\bibfnamefont {C.}~\bibnamefont
  {Miron}}, \bibinfo {author} {\bibfnamefont {M.}~\bibnamefont {Nagasono}},
  \bibinfo {author} {\bibfnamefont {T.}~\bibnamefont {Togashi}}, \bibinfo
  {author} {\bibfnamefont {M.}~\bibnamefont {Yabashi}}, \bibinfo {author}
  {\bibfnamefont {K.~L.}\ \bibnamefont {Ishikawa}}, \bibinfo {author}
  {\bibfnamefont {A.~K.}\ \bibnamefont {Kazansky}}, \bibinfo {author}
  {\bibfnamefont {N.~M.}\ \bibnamefont {Kabachnik}}, \ and\ \bibinfo {author}
  {\bibfnamefont {K.}~\bibnamefont {Ueda}},\ }\href {\doibase
  10.1103/PhysRevA.89.013415} {\bibfield  {journal} {\bibinfo  {journal} {Phys.
  Rev. A}\ }\textbf {\bibinfo {volume} {89}},\ \bibinfo {pages} {013415}
  (\bibinfo {year} {2014})}\BibitemShut {NoStop}%
\bibitem [{\citenamefont {Keldysh}(1965)}]{Keldysh65}%
  \BibitemOpen
  \bibfield  {author} {\bibinfo {author} {\bibfnamefont {L.}~\bibnamefont
  {Keldysh}},\ }\href@noop {} {\bibfield  {journal} {\bibinfo  {journal}
  {Soviet Physics JETP}\ }\textbf {\bibinfo {volume} {20}},\ \bibinfo {pages}
  {1307} (\bibinfo {year} {1965})}\BibitemShut {NoStop}%
\bibitem [{\citenamefont {Faisal}(1973)}]{Faisal1973}%
  \BibitemOpen
  \bibfield  {author} {\bibinfo {author} {\bibfnamefont {F.~H.~M.}\
  \bibnamefont {Faisal}},\ }\href {http://stacks.iop.org/0022-3700/6/i=4/a=011}
  {\bibfield  {journal} {\bibinfo  {journal} {Journal of Physics B: Atomic and
  Molecular Physics}\ }\textbf {\bibinfo {volume} {6}},\ \bibinfo {pages} {L89}
  (\bibinfo {year} {1973})}\BibitemShut {NoStop}%
\bibitem [{\citenamefont {Reiss}(1980)}]{Reiss1980}%
  \BibitemOpen
  \bibfield  {author} {\bibinfo {author} {\bibfnamefont {H.~R.}\ \bibnamefont
  {Reiss}},\ }\href {\doibase 10.1103/PhysRevA.22.1786} {\bibfield  {journal}
  {\bibinfo  {journal} {Phys. Rev. A}\ }\textbf {\bibinfo {volume} {22}},\
  \bibinfo {pages} {1786} (\bibinfo {year} {1980})}\BibitemShut {NoStop}%
\bibitem [{\citenamefont {Meyer}\ \emph {et~al.}(2006)\citenamefont {Meyer},
  \citenamefont {Cubaynes}, \citenamefont {O'Keeffe}, \citenamefont {Luna},
  \citenamefont {Yeates}, \citenamefont {Kennedy}, \citenamefont {Costello},
  \citenamefont {Orr}, \citenamefont {Ta\"{\i}eb}, \citenamefont {Maquet},
  \citenamefont {D\"usterer}, \citenamefont {Radcliffe}, \citenamefont
  {Redlin}, \citenamefont {Azima}, \citenamefont {Pl\"onjes},\ and\
  \citenamefont {Feldhaus}}]{Meyer2006}%
  \BibitemOpen
  \bibfield  {author} {\bibinfo {author} {\bibfnamefont {M.}~\bibnamefont
  {Meyer}}, \bibinfo {author} {\bibfnamefont {D.}~\bibnamefont {Cubaynes}},
  \bibinfo {author} {\bibfnamefont {P.}~\bibnamefont {O'Keeffe}}, \bibinfo
  {author} {\bibfnamefont {H.}~\bibnamefont {Luna}}, \bibinfo {author}
  {\bibfnamefont {P.}~\bibnamefont {Yeates}}, \bibinfo {author} {\bibfnamefont
  {E.~T.}\ \bibnamefont {Kennedy}}, \bibinfo {author} {\bibfnamefont {J.~T.}\
  \bibnamefont {Costello}}, \bibinfo {author} {\bibfnamefont {P.}~\bibnamefont
  {Orr}}, \bibinfo {author} {\bibfnamefont {R.}~\bibnamefont {Ta\"{\i}eb}},
  \bibinfo {author} {\bibfnamefont {A.}~\bibnamefont {Maquet}}, \bibinfo
  {author} {\bibfnamefont {S.}~\bibnamefont {D\"usterer}}, \bibinfo {author}
  {\bibfnamefont {P.}~\bibnamefont {Radcliffe}}, \bibinfo {author}
  {\bibfnamefont {H.}~\bibnamefont {Redlin}}, \bibinfo {author} {\bibfnamefont
  {A.}~\bibnamefont {Azima}}, \bibinfo {author} {\bibfnamefont
  {E.}~\bibnamefont {Pl\"onjes}}, \ and\ \bibinfo {author} {\bibfnamefont
  {J.}~\bibnamefont {Feldhaus}},\ }\href {\doibase 10.1103/PhysRevA.74.011401}
  {\bibfield  {journal} {\bibinfo  {journal} {Phys. Rev. A}\ }\textbf {\bibinfo
  {volume} {74}},\ \bibinfo {pages} {011401} (\bibinfo {year}
  {2006})}\BibitemShut {NoStop}%
\bibitem [{\citenamefont {Kazansky}\ and\ \citenamefont
  {Kabachnik}(2010)}]{Kazansky10a}%
  \BibitemOpen
  \bibfield  {author} {\bibinfo {author} {\bibfnamefont {A.~K.}\ \bibnamefont
  {Kazansky}}\ and\ \bibinfo {author} {\bibfnamefont {N.~M.}\ \bibnamefont
  {Kabachnik}},\ }\href {http://stacks.iop.org/0953-4075/43/i=3/a=035601}
  {\bibfield  {journal} {\bibinfo  {journal} {Journal of Physics B: Atomic,
  Molecular and Optical Physics}\ }\textbf {\bibinfo {volume} {43}},\ \bibinfo
  {pages} {035601} (\bibinfo {year} {2010})}\BibitemShut {NoStop}%
\bibitem [{\citenamefont {Kazansky}\ \emph {et~al.}(2010)\citenamefont
  {Kazansky}, \citenamefont {Sazhina},\ and\ \citenamefont
  {Kabachnik}}]{Kazansky10b}%
  \BibitemOpen
  \bibfield  {author} {\bibinfo {author} {\bibfnamefont {A.~K.}\ \bibnamefont
  {Kazansky}}, \bibinfo {author} {\bibfnamefont {I.~P.}\ \bibnamefont
  {Sazhina}}, \ and\ \bibinfo {author} {\bibfnamefont {N.~M.}\ \bibnamefont
  {Kabachnik}},\ }\href {\doibase 10.1103/PhysRevA.82.033420} {\bibfield
  {journal} {\bibinfo  {journal} {Phys. Rev. A}\ }\textbf {\bibinfo {volume}
  {82}},\ \bibinfo {pages} {033420} (\bibinfo {year} {2010})}\BibitemShut
  {NoStop}%
\bibitem [{\citenamefont {Bivona}\ \emph {et~al.}(2010)\citenamefont {Bivona},
  \citenamefont {Bonanno}, \citenamefont {Burlon},\ and\ \citenamefont
  {Leone}}]{Bivona10}%
  \BibitemOpen
  \bibfield  {author} {\bibinfo {author} {\bibfnamefont {S.}~\bibnamefont
  {Bivona}}, \bibinfo {author} {\bibfnamefont {G.}~\bibnamefont {Bonanno}},
  \bibinfo {author} {\bibfnamefont {R.}~\bibnamefont {Burlon}}, \ and\ \bibinfo
  {author} {\bibfnamefont {C.}~\bibnamefont {Leone}},\ }\href {\doibase
  10.1134/S1054660X10190023} {\bibfield  {journal} {\bibinfo  {journal} {Laser
  Physics}\ }\textbf {\bibinfo {volume} {20}},\ \bibinfo {pages} {2036}
  (\bibinfo {year} {2010})}\BibitemShut {NoStop}%
\bibitem [{\citenamefont {Gramajo}\ \emph {et~al.}(2016)\citenamefont
  {Gramajo}, \citenamefont {Della~Picca}, \citenamefont {Garibotti},\ and\
  \citenamefont {Arb\'o}}]{Gramajo16}%
  \BibitemOpen
  \bibfield  {author} {\bibinfo {author} {\bibfnamefont {A.~A.}\ \bibnamefont
  {Gramajo}}, \bibinfo {author} {\bibfnamefont {R.}~\bibnamefont
  {Della~Picca}}, \bibinfo {author} {\bibfnamefont {C.~R.}\ \bibnamefont
  {Garibotti}}, \ and\ \bibinfo {author} {\bibfnamefont {D.~G.}\ \bibnamefont
  {Arb\'o}},\ }\href {\doibase 10.1103/PhysRevA.94.053404} {\bibfield
  {journal} {\bibinfo  {journal} {Phys. Rev. A}\ }\textbf {\bibinfo {volume}
  {94}},\ \bibinfo {pages} {053404} (\bibinfo {year} {2016})}\BibitemShut
  {NoStop}%
\bibitem [{\citenamefont {Haber}\ \emph {et~al.}(2010)\citenamefont {Haber},
  \citenamefont {Doughty},\ and\ \citenamefont {Leone‡}}]{Haber2010}%
  \BibitemOpen
  \bibfield  {author} {\bibinfo {author} {\bibfnamefont {L.~H.}\ \bibnamefont
  {Haber}}, \bibinfo {author} {\bibfnamefont {B.}~\bibnamefont {Doughty}}, \
  and\ \bibinfo {author} {\bibfnamefont {S.~R.}\ \bibnamefont {Leone‡}},\
  }\href {\doibase 10.1080/00268976.2010.483133} {\bibfield  {journal}
  {\bibinfo  {journal} {Molecular Physics}\ }\textbf {\bibinfo {volume}
  {108}},\ \bibinfo {pages} {1241} (\bibinfo {year} {2010})},\ \Eprint
  {http://arxiv.org/abs/http://dx.doi.org/10.1080/00268976.2010.483133}
  {http://dx.doi.org/10.1080/00268976.2010.483133} \BibitemShut {NoStop}%
\bibitem [{\citenamefont {{Macri}}\ \emph {et~al.}(1998)\citenamefont
  {{Macri}}, \citenamefont {{Miraglia}}, \citenamefont {{Grabielle}},
  \citenamefont {{Colavecchia}}, \citenamefont {{Garibotti}},\ and\
  \citenamefont {{Gasaneo}}}]{Macri98}%
  \BibitemOpen
  \bibfield  {author} {\bibinfo {author} {\bibfnamefont {P.~A.}\ \bibnamefont
  {{Macri}}}, \bibinfo {author} {\bibfnamefont {J.~E.}\ \bibnamefont
  {{Miraglia}}}, \bibinfo {author} {\bibfnamefont {M.~S.}\ \bibnamefont
  {{Grabielle}}}, \bibinfo {author} {\bibfnamefont {F.~D.}\ \bibnamefont
  {{Colavecchia}}}, \bibinfo {author} {\bibfnamefont {C.~R.}\ \bibnamefont
  {{Garibotti}}}, \ and\ \bibinfo {author} {\bibfnamefont {G.}~\bibnamefont
  {{Gasaneo}}},\ }\href {\doibase 10.1103/PhysRevA.57.2223} {\bibfield
  {journal} {\bibinfo  {journal} {\pra}\ }\textbf {\bibinfo {volume} {57}},\
  \bibinfo {pages} {2223} (\bibinfo {year} {1998})}\BibitemShut {NoStop}%
\bibitem [{\citenamefont {{Arb{\'o}}}\ \emph {et~al.}(2008)\citenamefont
  {{Arb{\'o}}}, \citenamefont {{Miraglia}}, \citenamefont {{Gravielle}},
  \citenamefont {{Schiessl}}, \citenamefont {{Persson}},\ and\ \citenamefont
  {{Burgd{\"o}rfer}}}]{Arbo08a}%
  \BibitemOpen
  \bibfield  {author} {\bibinfo {author} {\bibfnamefont {D.~G.}\ \bibnamefont
  {{Arb{\'o}}}}, \bibinfo {author} {\bibfnamefont {J.~E.}\ \bibnamefont
  {{Miraglia}}}, \bibinfo {author} {\bibfnamefont {M.~S.}\ \bibnamefont
  {{Gravielle}}}, \bibinfo {author} {\bibfnamefont {K.}~\bibnamefont
  {{Schiessl}}}, \bibinfo {author} {\bibfnamefont {E.}~\bibnamefont
  {{Persson}}}, \ and\ \bibinfo {author} {\bibfnamefont {J.}~\bibnamefont
  {{Burgd{\"o}rfer}}},\ }\href {\doibase 10.1103/PhysRevA.77.013401} {\bibfield
   {journal} {\bibinfo  {journal} {\pra}\ }\textbf {\bibinfo {volume} {77}},\
  \bibinfo {eid} {013401} (\bibinfo {year} {2008})}\BibitemShut {NoStop}%
\bibitem [{\citenamefont {Wolkow}(1935)}]{Volkov}%
  \BibitemOpen
  \bibfield  {author} {\bibinfo {author} {\bibfnamefont {D.}~\bibnamefont
  {Wolkow}},\ }\href {\doibase 10.1007/BF01331022} {\bibfield  {journal}
  {\bibinfo  {journal} {Zeitschrift f\"ur Physik}\ }\textbf {\bibinfo {volume}
  {94}},\ \bibinfo {pages} {250} (\bibinfo {year} {1935})}\BibitemShut
  {NoStop}%
\bibitem [{\citenamefont {Nagele}\ \emph {et~al.}(2011)\citenamefont {Nagele},
  \citenamefont {Pazourek}, \citenamefont {Feist}, \citenamefont
  {Doblhoff-Dier}, \citenamefont {Lemell}, \citenamefont {Tok\'esi},\ and\
  \citenamefont {Burgd\"orfer}}]{Nagele11}%
  \BibitemOpen
  \bibfield  {author} {\bibinfo {author} {\bibfnamefont {S.}~\bibnamefont
  {Nagele}}, \bibinfo {author} {\bibfnamefont {R.}~\bibnamefont {Pazourek}},
  \bibinfo {author} {\bibfnamefont {J.}~\bibnamefont {Feist}}, \bibinfo
  {author} {\bibfnamefont {K.}~\bibnamefont {Doblhoff-Dier}}, \bibinfo {author}
  {\bibfnamefont {C.}~\bibnamefont {Lemell}}, \bibinfo {author} {\bibfnamefont
  {K.}~\bibnamefont {Tok\'esi}}, \ and\ \bibinfo {author} {\bibfnamefont
  {J.}~\bibnamefont {Burgd\"orfer}},\ }\href
  {http://stacks.iop.org/0953-4075/44/i=8/a=081001} {\bibfield  {journal}
  {\bibinfo  {journal} {Journal of Physics B: Atomic, Molecular and Optical
  Physics}\ }\textbf {\bibinfo {volume} {44}},\ \bibinfo {pages} {081001}
  (\bibinfo {year} {2011})}\BibitemShut {NoStop}%
\bibitem [{\citenamefont {Della~Picca}\ \emph {et~al.}(2013)\citenamefont
  {Della~Picca}, \citenamefont {Fiol},\ and\ \citenamefont
  {Fainstein}}]{DellaPicca13}%
  \BibitemOpen
  \bibfield  {author} {\bibinfo {author} {\bibfnamefont {R.}~\bibnamefont
  {Della~Picca}}, \bibinfo {author} {\bibfnamefont {J.}~\bibnamefont {Fiol}}, \
  and\ \bibinfo {author} {\bibfnamefont {P.~D.}\ \bibnamefont {Fainstein}},\
  }\href {http://stacks.iop.org/0953-4075/46/i=17/a=175603} {\bibfield
  {journal} {\bibinfo  {journal} {Journal of Physics B: Atomic, Molecular and
  Optical Physics}\ }\textbf {\bibinfo {volume} {46}},\ \bibinfo {pages}
  {175603} (\bibinfo {year} {2013})}\BibitemShut {NoStop}%
\bibitem [{\citenamefont {Chiril\ifmmode~\u{a}\else \u{a}\fi{}}\ and\
  \citenamefont {Potvliege}(2005)}]{Chirila05}%
  \BibitemOpen
  \bibfield  {author} {\bibinfo {author} {\bibfnamefont {C.~C.}\ \bibnamefont
  {Chiril\ifmmode~\u{a}\else \u{a}\fi{}}}\ and\ \bibinfo {author}
  {\bibfnamefont {R.~M.}\ \bibnamefont {Potvliege}},\ }\href {\doibase
  10.1103/PhysRevA.71.021402} {\bibfield  {journal} {\bibinfo  {journal} {Phys.
  Rev. A}\ }\textbf {\bibinfo {volume} {71}},\ \bibinfo {pages} {021402}
  (\bibinfo {year} {2005})}\BibitemShut {NoStop}%
\bibitem [{\citenamefont {Corkum}\ \emph {et~al.}(1994)\citenamefont {Corkum},
  \citenamefont {Burnett},\ and\ \citenamefont {Ivanov}}]{Corkum94}%
  \BibitemOpen
  \bibfield  {author} {\bibinfo {author} {\bibfnamefont {P.~B.}\ \bibnamefont
  {Corkum}}, \bibinfo {author} {\bibfnamefont {N.~H.}\ \bibnamefont {Burnett}},
  \ and\ \bibinfo {author} {\bibfnamefont {M.~Y.}\ \bibnamefont {Ivanov}},\
  }\href {\doibase 10.1364/OL.19.001870} {\bibfield  {journal} {\bibinfo
  {journal} {Opt. Lett.}\ }\textbf {\bibinfo {volume} {19}},\ \bibinfo {pages}
  {1870} (\bibinfo {year} {1994})}\BibitemShut {NoStop}%
\bibitem [{\citenamefont {Ivanov}\ \emph {et~al.}(1995)\citenamefont {Ivanov},
  \citenamefont {Corkum}, \citenamefont {Zuo},\ and\ \citenamefont
  {Bandrauk}}]{Ivanov95}%
  \BibitemOpen
  \bibfield  {author} {\bibinfo {author} {\bibfnamefont {M.}~\bibnamefont
  {Ivanov}}, \bibinfo {author} {\bibfnamefont {P.~B.}\ \bibnamefont {Corkum}},
  \bibinfo {author} {\bibfnamefont {T.}~\bibnamefont {Zuo}}, \ and\ \bibinfo
  {author} {\bibfnamefont {A.}~\bibnamefont {Bandrauk}},\ }\href {\doibase
  10.1103/PhysRevLett.74.2933} {\bibfield  {journal} {\bibinfo  {journal}
  {Phys. Rev. Lett.}\ }\textbf {\bibinfo {volume} {74}},\ \bibinfo {pages}
  {2933} (\bibinfo {year} {1995})}\BibitemShut {NoStop}%
\bibitem [{\citenamefont {Lewenstein}\ \emph {et~al.}(1995)\citenamefont
  {Lewenstein}, \citenamefont {Kulander}, \citenamefont {Schafer},\ and\
  \citenamefont {Bucksbaum}}]{Lewenstein95}%
  \BibitemOpen
  \bibfield  {author} {\bibinfo {author} {\bibfnamefont {M.}~\bibnamefont
  {Lewenstein}}, \bibinfo {author} {\bibfnamefont {K.~C.}\ \bibnamefont
  {Kulander}}, \bibinfo {author} {\bibfnamefont {K.~J.}\ \bibnamefont
  {Schafer}}, \ and\ \bibinfo {author} {\bibfnamefont {P.~H.}\ \bibnamefont
  {Bucksbaum}},\ }\href {\doibase 10.1103/PhysRevA.51.1495} {\bibfield
  {journal} {\bibinfo  {journal} {Phys. Rev. A}\ }\textbf {\bibinfo {volume}
  {51}},\ \bibinfo {pages} {1495} (\bibinfo {year} {1995})}\BibitemShut
  {NoStop}%
\bibitem [{\citenamefont {Arb\'o}\ \emph {et~al.}(2010)\citenamefont {Arb\'o},
  \citenamefont {Ishikawa}, \citenamefont {Schiessl}, \citenamefont {Persson},\
  and\ \citenamefont {Burgd\"orfer}}]{Arbo10a}%
  \BibitemOpen
  \bibfield  {author} {\bibinfo {author} {\bibfnamefont {D.~G.}\ \bibnamefont
  {Arb\'o}}, \bibinfo {author} {\bibfnamefont {K.~L.}\ \bibnamefont
  {Ishikawa}}, \bibinfo {author} {\bibfnamefont {K.}~\bibnamefont {Schiessl}},
  \bibinfo {author} {\bibfnamefont {E.}~\bibnamefont {Persson}}, \ and\
  \bibinfo {author} {\bibfnamefont {J.}~\bibnamefont {Burgd\"orfer}},\ }\href
  {\doibase 10.1103/PhysRevA.81.021403} {\bibfield  {journal} {\bibinfo
  {journal} {Phys. Rev. A}\ }\textbf {\bibinfo {volume} {81}},\ \bibinfo
  {pages} {021403} (\bibinfo {year} {2010})}\BibitemShut {NoStop}%
\bibitem [{\citenamefont {{Arb{\'o}}}\ \emph {et~al.}(2010)\citenamefont
  {{Arb{\'o}}}, \citenamefont {{Ishikawa}}, \citenamefont {{Schiessl}},
  \citenamefont {{Persson}},\ and\ \citenamefont {{Burgd{\"o}rfer}}}]{Arbo10b}%
  \BibitemOpen
  \bibfield  {author} {\bibinfo {author} {\bibfnamefont {D.~G.}\ \bibnamefont
  {{Arb{\'o}}}}, \bibinfo {author} {\bibfnamefont {K.~L.}\ \bibnamefont
  {{Ishikawa}}}, \bibinfo {author} {\bibfnamefont {K.}~\bibnamefont
  {{Schiessl}}}, \bibinfo {author} {\bibfnamefont {E.}~\bibnamefont
  {{Persson}}}, \ and\ \bibinfo {author} {\bibfnamefont {J.}~\bibnamefont
  {{Burgd{\"o}rfer}}},\ }\href {\doibase 10.1103/PhysRevA.82.043426} {\bibfield
   {journal} {\bibinfo  {journal} {\pra}\ }\textbf {\bibinfo {volume} {82}},\
  \bibinfo {eid} {043426} (\bibinfo {year} {2010})}\BibitemShut {NoStop}%
\bibitem [{\citenamefont {Tong}\ and\ \citenamefont {Chu}(1997)}]{Tong97}%
  \BibitemOpen
  \bibfield  {author} {\bibinfo {author} {\bibfnamefont {X.~M.}\ \bibnamefont
  {Tong}}\ and\ \bibinfo {author} {\bibfnamefont {S.~I.}\ \bibnamefont {Chu}},\
  }\href {http://dx.doi.org/10.1016/S0301-0104(97)00063-3} {\bibfield
  {journal} {\bibinfo  {journal} {Chem. Phys.}\ }\textbf {\bibinfo {volume}
  {217}},\ \bibinfo {pages} {119} (\bibinfo {year} {1997})}\BibitemShut
  {NoStop}%
\bibitem [{\citenamefont {Tong}\ and\ \citenamefont {Chu}(2000)}]{Tong00}%
  \BibitemOpen
  \bibfield  {author} {\bibinfo {author} {\bibfnamefont {X.-M.}\ \bibnamefont
  {Tong}}\ and\ \bibinfo {author} {\bibfnamefont {S.-I.}\ \bibnamefont {Chu}},\
  }\href {\doibase 10.1103/PhysRevA.61.031401} {\bibfield  {journal} {\bibinfo
  {journal} {Phys. Rev. A}\ }\textbf {\bibinfo {volume} {61}},\ \bibinfo
  {pages} {031401} (\bibinfo {year} {2000})}\BibitemShut {NoStop}%
\bibitem [{\citenamefont {Tong}\ and\ \citenamefont {Lin}(2005)}]{Tong05}%
  \BibitemOpen
  \bibfield  {author} {\bibinfo {author} {\bibfnamefont {X.~M.}\ \bibnamefont
  {Tong}}\ and\ \bibinfo {author} {\bibfnamefont {C.~D.}\ \bibnamefont {Lin}},\
  }\href {http://stacks.iop.org/0953-4075/38/i=15/a=001} {\bibfield  {journal}
  {\bibinfo  {journal} {Journal of Physics B: Atomic, Molecular and Optical
  Physics}\ }\textbf {\bibinfo {volume} {38}},\ \bibinfo {pages} {2593}
  (\bibinfo {year} {2005})}\BibitemShut {NoStop}%
\bibitem [{Note1()}]{Note1}%
  \BibitemOpen
  \bibinfo {note} {Here the $2\pi $ equivalence $a \equiv b $ means that $(a
  -b)/2\pi $ is integer and $0\leq \phi < 2\pi $.}\BibitemShut {Stop}%
\end{thebibliography}%

\end{document}